\documentclass[fleqn,usenatbib,onecolumn]{mnras}

\usepackage{mathptmx}

\usepackage[T1]{fontenc}

\DeclareRobustCommand{\VAN}[3]{#2}
\let\VANthebibliography\thebibliography
\def\thebibliography{\DeclareRobustCommand{\VAN}[3]{##3}\VANthebibliography}

\usepackage{graphicx}	
\usepackage{amsmath}	
\usepackage{amssymb}	
\usepackage{epsfig}
\usepackage{dcolumn}
\usepackage{paralist}
\usepackage{comment}
\usepackage{graphicx}
\usepackage{
multicol}
\usepackage{multirow}
\usepackage{color,soul}
\usepackage{suffix}
\usepackage{mathtools}
\usepackage{wrapfig}

\title[Mass distribution of a gravitational lens]{Recovering the mass distribution of an extended gravitational lens}

\author[S. G. Turyshev and V. T. Toth]{
Slava G. Turyshev$^{1}$, Viktor T. Toth$^2$
\\
$^1$Jet Propulsion Laboratory, California Institute of Technology,
4800 Oak Grove Drive, Pasadena, CA 91109-0899, USA\\
$^2$Ottawa, Ontario K1N 9H5, Canada
}

\date{Accepted 2022 April 22. Received 2022 April 22; in original form 2022 March 04}

\pubyear{2022}

\begin{document}
\label{firstpage}
\pagerange{\pageref{firstpage}--\pageref{lastpage}}
\maketitle

\begin{abstract}
We investigate the possibility of determining the mass distribution of a gravitational lens via lensing observations. We consider an extended, compact gravitational lens, representing its static external gravitational potential via an infinite set of symmetric trace free (STF) multipole moments. Within the wave-optical treatment, we evaluate the caustics formed in the lens's point spread function (PSF). We study the only quantity that is available in astronomical lensing observations: the image of that PSF formed by an imaging telescope. This observable may be used to recover some physical characteristics of the lens, including its shape, orientation and composition. Illustrating this, we study exotic gravitational lenses formed by several well-known solids with uniform density. We show that when moments beyond the quadrupole are observed, some of the symmetry properties of the lens can be recovered. The presence of an octupole moment implies breaking the ``north-south'' symmetry of the mass distribution in the lens. The presence of a rotated hexadecapole moment implies breaking axial symmetry. As such, if observations of lensed images allow the reconstruction of these moments, important information about the mass distribution and dynamics of the lens can be obtained. This may help with choosing the most appropriate mass profile that is used to characterize the mass distribution of astrophysical lenses, such as the dark matter halos that are presumed to contain most of the mass of galaxies and clusters of galaxies. Our results are novel and offer new insight into gravitational lensing by realistic astrophysical systems.

\end{abstract}




\section{Introduction}

Gravitational lensing can reveal information not only about the object being lensed but about the mass distribution of the lens itself. However, it is known that realistic gravitational lenses are complex and are difficult to model analytically. A typical approach  involves perturbing the spherically symmetric Schwarzschild lens by a quadrupole moment, parameterized by a simple mass model (see discussion in \cite{Schneider-Ehlers-Falco:1992}). Although we use gravitational lensing observations to make important conjectures about mass distributions in the universe, the mass models that are typically employed for this purpose  (e.g., \cite{Keeton2001}) are rather too simplistic and often poorly describe astrophysical reality. The perturbative approach cannot easily capture more subtle features of the mass distribution of the lens, and a reliable wave-theoretical reconstruction of the actual, observed lens images is not always possible. Modeling the contributions of the higher moments is even less intuitive and is usually done relying on  semianalytical or entirely numerical analysis.

Alternatively, gravitational potentials of compact extended lenses may be expressed in terms of spherical harmonics and related multipole moments of the external gravitational field of the lens (see discussion \cite{Turyshev-Toth:2021-multipoles}).
To explore astrophysical lenses in the most general case, we recently developed a new approach to study extended gravitational lenses \cite{Turyshev-Toth:2021-multipoles,Turyshev-Toth:2021-STF-moments}. For that, we considered the propagation of high-frequency electromagnetic (EM) waves in the vicinity of an extended gravitating body. Using the Mie theory \cite{Mie:1908,Born-Wolf:1999}, we solved the Maxwell equations on the background of a static gravitational field, while working within the first post-Newtonian approximation of the general theory of relativity.

The new solution describes the EM field deposited on the image plane located in various regions of interest, including those of strong and weak interference and that of geometric optics. We have shown that deviations from spherical symmetry in the lensing object's gravitational field is evident only in the strong interference region, where it leads to caustics of various orders appearing in the lens's point spread function (PSF) \cite{Turyshev-Toth:2021-multipoles,Turyshev-Toth:2021-caustics}. In the other two regions, the optical properties of the lens are consistent with those of a monopole lens \cite{Turyshev-Toth:2017,Turyshev-Toth:2019-extend,Turyshev-Toth:2020-image,Turyshev-Toth:2020-extend,Toth-Turyshev:2020,Turyshev-Toth:2021-all-regions}.  Thus, to capture the most interesting behavior in lensing by a body with arbitrary mass distribution, we need to consider the strong interference region.

Further generalizing the newly developed wave-optical treatment, we extended the description of gravitational lensing to a generic mass distribution \cite{Turyshev-Toth:2021-STF-moments}. For that, we modeled the external static gravitational field of an extended object in the most general case,  taking the potential in the form of an infinite series of symmetric trace-free (STF) mass multipole moments. Such a representation of the gravitational potential in terms of the STF Cartesian tensors is equivalent to that expressed in the form of spherical harmonics. The advantage of using the STF formalism is that it allows us to derive the gravitational phase shift for arbitrary mass distributions, not restricted to, e.g., axial symmetry. This generalizes our previous results \cite{Turyshev-Toth:2021-multipoles,Turyshev-Toth:2021-imaging,Turyshev-Toth:2021-all-regions,Turyshev-Toth:2021-quartic}. Using our results, we are able to model not only the caustics associated with the PSF, but also convolve the result with the PSF of an imaging telescope, leading to an accurate wave-theoretical model of the image seen by such an instrument. In short, we can model the Einstein ring, Einstein cross, or more complex lensed images accurately, with an appropriate wave-optical treatment.

In Ref.~\cite{Turyshev-Toth:2021-STF-moments}, we have shown that at each STF multipole order, only two parameters are required to describe the effect of an extended lens. This is simpler than expected and it applies even to objects without any symmetries.  Although it is common to account only for the lens plane components of the lensing potential (see \cite{Schneider-Ehlers-Falco:1992} for details), we were able to develop insight for such a thin lens approximation working rigorously from the first principles.  That result suggested that observations from a single vantage point are limited to only two combinations of the transverse-traceless STF tensor moments of a gravity field, thus precluding reconstruction of the full 3-dimensional mass distribution. Nonetheless, if the parameters of the projection can be determined with some accuracy, important information of potential astrophysical significance can be obtained about the lens.

Recognizing the value of this development, there is a need to consider its possible practical applications. Some of the important questions include: What is the number of moments needed to achieve the best modeling accuracy? To what extent is it possible to determine the shape and distribution of matter within the lens from examining images produced by it? Our paper aims to provide guidance by presenting specific, idealized examples of gravitational lenses and investigating some of their properties. The choice of the bodies emphasizes the fact that results obtained here are generic and are valid for any extended body with arbitrary sets of multipole moments, as discussed in \cite{Turyshev-Toth:2021-STF-moments}.

This paper is organized as follows:
In Section \ref{sec:opt-prop} we consider gravitational lensing by extended compact bodies with arbitrary mass distributions while expressing their external gravitational potentials via infinite sets of the STF tensor multipole moments. We summarize a wave-optical solution that was obtained in \cite{Turyshev-Toth:2021-STF-moments} to describe light propagation under such conditions. In Section \ref{sec:lens-Plat} we study lensing by several simple geometric shapes that are treated as gravitational lenses. Specifically we consider lensing by a sphere, a cylinder, a right circular cone, an ellipsoid, a cuboid and a trirectangular tetrahedron. In Section~\ref{sec:appl} we show how progressively including higher moments leads to being able to recover important symmetry properties of the mass distribution of the extended lensing object. In Section~\ref{sec:end} we discuss results and outline the next steps in our investigation.

\section{Optical properties of an extended lens}
\label{sec:opt-prop}

We consider an isolated extended object acting as a gravitational lens \cite{Turyshev-Toth:2021-STF-moments}. To characterize the gravitational field of a generic lens, following \cite{Turyshev-Toth:2017,Turyshev-Toth:2021-multipoles},  we use a static harmonic metric in the first post-Newtonian approximation of the general theory of relativity.  The line element for this metric in lens-centric spherical coordinates $(r,\theta,\phi)$, to the accuracy sufficient to describe light propagation in a weak gravitational field \cite{Turyshev-Toth:2013}, may be given as
\begin{eqnarray}
ds^2&=&\Big(1+c^{-2}U+{\cal O}(c^{-4})\Big)^{-2}c^2dt^2-
\Big(1+c^{-2}U+{\cal O}(c^{-4})\Big)^2\big(dr^2+r^2\big(d\theta^2+\sin^2\theta d\phi^2\big)\big),~~~
\label{eq:metric-gen}
\end{eqnarray}
where the Newtonian potential, $U$, generated by the mass density $\rho({\vec r})$ characterizing the source, is  given as usual:
\begin{eqnarray}
U({\vec r})=G\int\frac{\rho({\vec r}')d^3{\vec r}'}{|{\vec r}-{\vec r}'|}.
\label{eq:w-PN}
\end{eqnarray}

With the preliminaries above, we study the propagation of a high-frequency plane EM wave (i.e., neglecting terms $\propto(kr)^{-1}$, where $k=2\pi/\lambda$ is the wavenumber and $\lambda$ is the wavelength) in the vicinity of the lens.  The lens's Schwarzschild radius is  $r_g=2GM/c^2$, where $M$ is its mass. We assume that the wave is emitted by a point source, which is located at a large distance $r_0$ from the lens, so $r_g/r_0\ll 1$. We consider that this field is observed in an image plane also located at a large distance $r$ from the lens, such that $r_g/r\ll1$.

\subsection{The EM field on the image plane}

Following closely the notations in \cite{Turyshev-Toth:2021-STF-moments}, we represent the trajectory of an incident light ray as
{}
\begin{eqnarray}
\vec{r}(t)&=&\vec{r}_{0}+\vec{k}c(t-t_0)+{\cal O}(r_g),
\label{eq:x-Newt0}
\end{eqnarray}
where $\vec k$ is a unit wave vector in the direction of propagation of the incident light ray and $\vec r_0$ represents the source position. We use ${\vec b}=[[{\vec k}\times{\vec r}_0]\times{\vec k}]$ as the vector impact parameter corresponding to the unperturbed light ray's trajectory.
We use the affine parameter $\tau=\tau(t)$ to characterize the light ray's path (see details in Appendix~B in \cite{Turyshev-Toth:2017}):
{}
\begin{eqnarray}
\tau &=&({\vec k}\cdot {\vec x})=({\vec k}\cdot {\vec r}_{0})+c(t-t_0).
\label{eq:x-Newt*=0}
\end{eqnarray}
We use a lens-centric cylindrical coordinate system $(\rho,\phi,z)$ with its $z$-coordinate oriented along the wavevector $\vec k$, a unit vector in the unperturbed direction of the propagation of the incident wave; thus, we have $\tau=z$. The value of $\tau$ starts as negative at the originating point of the light ray ($\alpha\simeq\pi$), reaches zero at the point of closest approach to the lens (characterized by $\alpha=\pi/2$) and progresses through increasing positive values after departing the vicinity of the lens ($\alpha\simeq 0$). With the help of this parameter, we can rewrite (\ref{eq:x-Newt0}) as
{}
\begin{eqnarray}
{\vec r}(\tau)&=&{\vec b}+{\vec k} \tau+{\cal O}(r_g),
\qquad {\rm with} \qquad
 ||{\vec r}(\tau)|| \equiv r(\tau) =\sqrt{b^2+\tau^2}+{\cal O}(r_g).
\label{eq:b0}
\end{eqnarray}

We also introduce a light ray's impact parameter, $\vec b$, and coordinates on the image plane, $\vec x$, located in the strong interference region of the lens at distance $z$ from the lens \cite{Turyshev-Toth:2017,Turyshev-Toth:2021-STF-moments}.  Thus, we then have:
{}
\begin{eqnarray}
\vec k&=&(0,0,1),
\label{eq:note-k}\\
{\vec b}&=&b(\cos\phi_\xi,\sin \phi_\xi,0)=b\, \vec m,
\label{eq:note-b}\\
{\vec x}&=&\rho(\cos\phi,\sin \phi,0)=\rho \, \vec n.
\label{eq:note-x}
\end{eqnarray}

Using this parametrization, we solved the gravitational Mie problem (see discussion in \cite{Born-Wolf:1999,Turyshev-Toth:2017}) to the required order ($\rho\lesssim r_g\ll r$) and found that the EM field is given by \cite{Turyshev-Toth:2021-multipoles,Turyshev-Toth:2021-all-regions}:
{}
\begin{eqnarray}
    \left( \begin{aligned}
{E}_\rho& \\
{H}_\rho& \\
  \end{aligned} \right) =\left( \begin{aligned}
{H}_\phi& \\
-{E}_\phi& \\
  \end{aligned} \right) &=&
\frac{E_0}{r_0}e^{i\Omega(t)}
A(\vec x)
 \left( \begin{aligned}
 \cos\phi& \\
 \sin\phi& \\
  \end{aligned} \right)+{\cal O}(r_g^2,\rho^2/z^2),
  \label{eq:DB-sol-rho}
\end{eqnarray}
 where   $\Omega(t)=\big(k(r_0+r)-\omega t\big)$. The remaining components of the EM field are small,  $({E}_z, {H}_z)= {\cal O}({\rho}/{z})$.

The amplification factor of the EM field, $A(\rho,\phi)\equiv A(\vec x)$, is given as
{}
\begin{eqnarray}
A(\vec x) =
 \frac{k}{ir}\frac{1}{2\pi}\int d^2\vec b \,\exp\Big[ik\Big(\frac{1}{2 r}({\vec b} - \vec x)^2+
 \frac{2}{c^2}\int^{\tau}_{\tau_0}  U({\vec b},\tau') d\tau'
 \Big)\Big],
  \label{eq:amp-A}
\end{eqnarray}
where the phase of the integral of (\ref{eq:amp-A}) is known as the Fermat potential of gravitational lensing \cite{Schneider-Ehlers-Falco:1992}. The first term in the phase of (\ref{eq:amp-A}) is the phase shift associated with the geometric delay. The remainder of the expression represents the gravitational delay accumulated by the EM wave as it travels through the phase shift accumulated by EM wave as it travels from the source to the image plane on the background on the gravitational potential $U$.

\subsection{Computing the eikonal phase for a generic gravitational field}
\label{sec:eik-phase-axsym}

Considering a generic case, it was shown \cite{Thorne:1980,Blanchet-Damour:1986,Blanchet-Damour:1989,Kopeikin:1997,Mathis-LePoncinLafitte:2007,Soffel-Han:2019} that the scalar gravitational potential  (\ref{eq:w-PN}) may be given equivalently in the following form:
{}
\begin{eqnarray}
U(\vec r)&=& GM\sum_{\ell\geq 0}\frac{(2\ell-1)!!}{\ell !}{\cal T}_L\frac{\hat n_L}{r^{\ell+1}},
\label{eq:pot_w_0STF}
\end{eqnarray}
where $r=|{\vec r}|$, $M$ is the mass of the body and ${\cal T}_L\equiv {\cal T}^{<a_1...a_\ell>}$ are the body's normalized Newtonian STF mass multipole moments, given as
{}
\begin{eqnarray}
M&=&\int d^3{\vec r'}\, \rho({\vec r'}),\qquad
{\cal T}^{<a_1...a_\ell>}=
\int d^3{\vec x'}\, \rho({\vec r'})\, x'{}^{<a_1...a_\ell>},
\label{eq:mom}
\end{eqnarray}
where $x^{<a_1...a_\ell>}=x^{<a_1}x^{a_2...}x^{a_\ell>}\equiv \hat x^L$, while the angle  brackets $<...>$ and $\hat{x}$ denote the STF operator \cite{Hamermesh1962}, also $x^a$ here is $a$-th component of the 3-dimensional vector with its unit vector defined as usual $n^a=x^a/r$, thus $\hat n_L$ is the STF combination of $\ell$ unit vectors $n^{<a_1...a_\ell>}\equiv \hat n_L$. Without loss of generality, we set the origin of the coordinate system at the body's center-of-mass, which allows us to eliminate the dipole moment ${\cal T}^a$ from the expansion (\ref{eq:pot_w_0STF}).

The first few terms of (\ref{eq:pot_w_0STF}) are given as
{}
\begin{eqnarray}
U(\vec r)&=&G\Big\{\frac{M}{r}+ \frac{3{\cal T}^{<ij>}}{2r^5}x^ix^j +\frac{5{\cal T}^{<ijk>}}{2r^7}x^ix^jx^k+\frac{35{\cal T}^{<ijkl>}}{8r^9}x^ix^jx^kx^l+{\cal O}(r^{-6})\Big\}.
\label{eq:pot_w_0STF2}
\end{eqnarray}
This Cartesian multipole expansion of the Newtonian gravitational potential (\ref{eq:pot_w_0STF})--(\ref{eq:pot_w_0STF2}) is equivalent to expansion in terms of spherical harmonics  (e.g., see discussion in \cite{Turyshev-Toth:2021-STF-moments}). However, the use of the STF mass moment tensors simplifies the task of solving the light propagation equations in the post-Newtonian formalism of the general theory of  relativity \cite{Kopeikin:1997,Mathis-LePoncinLafitte:2007,Soffel-Han:2019}.

Using the light trajectory parametrization $\vec r=\vec r(\vec b,\tau)$ from (\ref{eq:b0}), we obtain the following expression for the gravitational phase shift (i.e., the second term in the phase of (\ref{eq:amp-A}), see derivation details in  \cite{Turyshev-Toth:2021-STF-moments}):
{}
\begin{eqnarray}
\varphi(\vec b) &=&\frac{2k}{c^2}\int^{\tau}_{\tau_0}  U({\vec b},\tau') d\tau'=
 kr_g\ln 4k^2rr_0 -2kr_g\Big(\ln kb-\sum_{\ell=2}^\infty
\frac{(2\ell-2)!!}{\ell! \,b^\ell} \sqrt{t^{+2}_\ell +t^{\times2}_\ell}\cos[\ell(\phi_\xi-\phi_\ell)]  \Big)+{\cal O}(r_g^2),~~~~
\label{eq:eik-ph-shift_tpm}
\end{eqnarray}
where $t^+_\ell$ and $t^\times_\ell$ are the transverse trace-free (TT) components of the  ${\cal T}^{<a_1....a_\ell>}$ tensor and the angle $\phi_\ell$ is given by
{}
\begin{eqnarray}
\cos[\ell\phi_\ell]=\frac{t^+_\ell}{\sqrt{t^{+2}_\ell +t^{\times2}_\ell}},\qquad
\sin[\ell\phi_\ell]=\frac{t^\times_\ell}{\sqrt{t^{+2}_\ell +t^{\times2}_\ell}}.
  \label{eq:tt52a7}
\end{eqnarray}
Note that the TT operation is understood with respect to the direction of the wave-vector $\vec k$ (\ref{eq:note-k}). Thus, any TT-projected quantity will be in the plane set by the vector of the impact parameter  (\ref{eq:note-b})  (see discussion in \cite{Turyshev-Toth:2021-STF-moments}).

In \cite{Turyshev-Toth:2021-STF-moments}, we computed  several low order terms in (\ref{eq:eik-ph-shift_tpm}), namely for $\ell=2,3,4$. We use parameterizations for the vectors $\vec k$ and $\vec m$ as given by (\ref{eq:note-k})--(\ref{eq:note-b}).  Thus, the lowest order $t^+_\ell$ and $t^\times_\ell$ are given as
{}
\begin{eqnarray}
t^+_2&=&{\textstyle\frac{1}{2}} ({\cal T}_{11}-{\cal T}_{22}), \qquad\qquad\qquad\quad~~
t^\times_2 ~~=~~ {\cal T}_{12},
\label{eq:eik-tt2}\\
t^+_3&=&{\textstyle\frac{1}{4}}({\cal T}_{111}-3{\cal T}_{122}),
\qquad\qquad\qquad~
 t^\times_3 ~~=~~ {\textstyle\frac{1}{4}} (3{\cal T}_{112}-{\cal T}_{222}),
\label{eq:eik-tt3}\\
t^+_4&=&{\textstyle\frac{1}{8}}({\cal T}_{1111}+{\cal T}_{2222}-6{\cal T}_{1122}) , \qquad  \,  t^\times_4 ~~=~~ {\textstyle\frac{1}{2}} ({\cal T}_{1112}-{\cal T}_{1222}).
\label{eq:eik-tt4}
\end{eqnarray}

We observe that at each order, the gravitational phase shift is determined by just the two degrees of freedom of the corresponding TT-projected STF multipole moment, $t^{+}_\ell $ and $t^{\times}_\ell$. In other words, at each STF order $\ell$, the amplitude $(t^{+2}_\ell +t^{\times2}_\ell)^\frac{1}{2}$,  and the rotation angle $\ell\phi_\ell$ of the gravitational phase shift (\ref{eq:eik-ph-shift_tpm})--(\ref{eq:tt52a7}) are set by only two combinations of the TT-projected STF mass multipole moments, $t^{+}_\ell$ and $t^{\times}_\ell$ (see details in \cite{Turyshev-Toth:2021-STF-moments}). We observe that the longitudinal components of the gravitational potential (i.e., those that are orthogonal to the line of sight direction taken to be along the $\vec k$ vector) of the lensing mass distribution are not accessible. Through astronomical observations we observe only the TT-projected STF multipole moments of any mass distribution.

\subsection{Optical properties of the extended lens}

At this point we have all the necessary ingredients to consider imaging of point sources with extended lenses. This is important, as such images explicitly reveal the structure of the gravitational lens by producing a caustic of a particular shape \cite{Turyshev-Toth:2021-multipoles,Turyshev-Toth:2021-caustics} as opposed to imaging of extended sources that will result in Einstein rings \cite{Turyshev-Toth:2020-extend}.

\subsubsection{Image formation of point sources with extended lenses}

To consider the image formation properties of an extended lens we need to establish its PSF. We do that by substituting the results (\ref{eq:eik-ph-shift_tpm})--(\ref{eq:tt52a7})  in (\ref{eq:amp-A}). We get
{}
\begin{eqnarray}
A(\vec x) ~~=~~ e^{ikr_g\ln 4k^2rr_0}
 \frac{k}{ir}\frac{1}{2\pi}\iint d^2\vec b \,\exp\Big[ik\Big(\frac{1}{2 \tilde r}({\vec b} - \vec x)^2-
 2r_g\Big(\ln kb-\sum_{\ell=2}^\infty
\frac{(2\ell-2)!!}{\ell! \,b^\ell} \sqrt{t^{+2}_\ell +t^{\times2}_\ell}\cos[\ell(\phi_\xi-\phi_\ell)] \Big)\Big)\Big].
  \label{eq:amp-A+}
\end{eqnarray}

In general, this integral must be treated numerically. However, there are two important observations:
\begin{inparaenum}[1)]
\item As the contribution of the $\ell$-th multipole moment scales as $1/b^\ell$, at some distance from the lens, the overall lensing potential approaches that of a monopole.
\item For a weakly aspherical lens, multipole moments  are small, making it possible to evaluate (\ref{eq:amp-A+}) using the method of stationary phase with respect to the radial variable, $b$, as we did in \cite{Turyshev-Toth:2021-multipoles,Turyshev-Toth:2021-all-regions}.
\end{inparaenum}
Specifically, we express the integration variables in the double integral (\ref{eq:amp-A+}) using the polar coordinates $(b,\phi_\xi)$ and evaluate the radial integral from a finite value $R$ that characterizes the extent of the lens. Essentially, this means that we treat the lens as an opaque object, considering only light with impact parameter $b>R$.

Under the conditions summarized above, we found that we can evaluate the radial integral in (\ref{eq:amp-A+}) using the method of stationary phase (see \cite{Turyshev-Toth:2017,Turyshev-Toth:2021-multipoles,Turyshev-Toth:2021-all-regions}), which leads to the following form for the amplification factor:
{}
\begin{eqnarray}
A(\vec x) ~~=~~
\sqrt{2\pi kr_g}e^{i\sigma_0}e^{ik(r_0+r+r_g\ln 4k^2rr_0)}B(\vec x),
  \label{eq:amp-A2-STF}
\end{eqnarray}
where  $B(\vec x)$ is the  generalized complex amplitude of the EM field in case of an arbitrary, weakly aspherical lens:
{}
\begin{eqnarray}
B(\vec x) ~~=~~
\frac{1}{2\pi}\int_0^{2\pi} d\phi_\xi \exp\Big[-ik\Big(\sqrt{\frac{2r_g}{\tilde r}}\rho\cos(\phi_\xi-\phi)-
2r_g\sum_{\ell=2}^\infty
\frac{(2\ell-2)!!}{\ell! \,(\sqrt{2r_g\tilde r})^\ell} \sqrt{t^{+2}_\ell +t^{\times2}_\ell}\cos[\ell(\phi_\xi-\phi_\ell)]\Big)\Big].
  \label{eq:B2-STF}
\end{eqnarray}

Using (\ref{eq:B2-STF}), we form the PSF of a generic lens (see details in \cite{Turyshev-Toth:2021-multipoles,Turyshev-Toth:2021-STF-moments}) that is given as the square of the complex amplitude, namely
{}
\begin{eqnarray}
{\rm PSF}({\vec x})&=&|B({\vec x})|^2.
 \label{eq:psf=}
\end{eqnarray}
This PSF can be used for the practical modeling of gravitational lenses, especially for imaging of faint sources \cite{Turyshev-Toth:2021-imaging}.

In \cite{Turyshev-Toth:2021-multipoles,Turyshev-Toth:2021-caustics} we considered the PSFs formed in the presence of various multiples of an axisymmetric lens. Result (\ref{eq:psf=}) with $B(\vec x)$ from (\ref{eq:B2-STF}) generalizes it to an arbitrary mass distribution. We can see that at each order $\ell$ the caustic formed on the image plane will be characterized by only two parameters: its magnitude $Q_\ell=\frac{(2\ell-2)!!}{\ell! \,({2r_g\tilde r})^{\ell/2}} \big({t^{+2}_\ell +t^{\times2}_\ell}\big)^\frac{1}{2}$ and the rotation angle $\phi_\ell$ given by (\ref{eq:tt52a7}), thus resembling the case of axisymmetric lenses \cite{Turyshev-Toth:2021-multipoles,Turyshev-Toth:2021-caustics}. Below we consider the implications of such a simplification on lensing observations.

\subsubsection{Observing images of point sources with extended lenses}

As it is known \cite{Turyshev-Toth:2021-imaging}, the PSF is the image of a point source that is formed by an extended lens. For any extended lens with deviations from spherical symmetry such an image will come in the form of a combination of various caustics that represent various multipolar deformations \cite{Turyshev-Toth:2021-multipoles,Turyshev-Toth:2021-caustics}. However, in typical astronomical observations, the caustics are not directly observed. Astronomical telescopes are used to look at the lens, instead of studying the potentially very large image projected by the distant gravitational lens in the image plane.

In practice, in astronomical observations a telescope is usually positioned inside the caustic region formed on the image plane. The telescope looks back toward the lens and sees either Einstein ring (i.e., $\ell=0$, if the lens is spherically-symmetric) or Einstein cross (i.e., $\ell=2$, if, in addition to a monopole, a small quadrupole moment is present) or other, more elaborate petal structures characteristic of multipoles of higher order $\ell$ (see discussion in \cite{Turyshev-Toth:2021-caustics} and Fig.~\ref{fig:montage}). Here we summarize the tools developed to describe such observations.

With the knowledge of the PSF of the extended gravitational lens, we consider the EM field as it is seen through an imaging telescope. To do this, we treat the imaging telescope as a thin lens and perform a Fourier transform of the EM field (\ref{eq:DB-sol-rho}) characterized by the complex amplitude $A(\vec x)$, from (\ref{eq:amp-A2-STF})--(\ref{eq:B2-STF}).  For that, we use the standard approach (e.g.,  \cite{Born-Wolf:1999,Goodman:2017}; see also details on the specific application in \cite{Turyshev-Toth:2020-photom,Turyshev-Toth:2020-image,Turyshev-Toth:2021-imaging}), and we introduce  ${\vec x}_i$, representing a point on the focal plane of the optical telescope:
{}
\begin{eqnarray}
 \{{\vec x}_i\}&\equiv& (x_i,y_i,0)=\rho_i\big(\cos\phi_i,\sin\phi_i,0\big).
 \label{eq:coord}
\end{eqnarray}

Following \cite{Turyshev-Toth:2021-imaging,Turyshev-Toth:2021-STF-moments}, we obtain the amplification factor, $I({\vec x},{\vec x}_i)$  of the optical system consisting of the lens and the imaging telescope (i.e., the convolution of the PSF of a lens with that of an optical telescope), that in the  case of a generic extended lens with arbitrary symmetry takes the form
 {}
\begin{eqnarray}
I({\vec x},{\vec x}_i)=|{\cal A}({\vec x},{\vec x}_i)|^2,
  \label{eq:Pv}
\end{eqnarray}
where $I({\vec x},{\vec x}_i)$ is the intensity distribution corresponding to the image of a point source as seen by the imaging telescope (see  \cite{Turyshev-Toth:2021-imaging} for details) and  ${\cal A}({\vec x},{\vec x}_i)$ is the normalized Fourier transform of the  amplitude $B(\vec x)$ from  (\ref{eq:B2-STF}):
{}
\begin{eqnarray}
{\cal A}({\vec x},{\vec x}_i) ~=~
\frac{1}{2\pi}\int_0^{2\pi} d\phi_\xi \,
  \Big(\frac{
2J_1(u(\phi_\xi,\phi_i)\frac{1}{2}d)}{u(\phi_\xi,\phi_i) \frac{1}{2}d}\Big)
\exp\Big[-ik\Big(\sqrt{\frac{2r_g}{ r}} \rho\cos(\phi_\xi-\phi)-
2r_g\sum_{\ell=2}^\infty
\frac{(2\ell-2)!!}{\ell! \,(\sqrt{2r_g\tilde r})^\ell} \sqrt{t^{+2}_\ell +t^{\times2}_\ell}\cos[\ell(\phi_\xi-\phi_\ell)]\Big)\Big],
  \label{eq:BinscER}
\end{eqnarray}
with $d$ being the telescope's aperture and $u(\phi_\xi,\phi_i)$ is given by
{}
\begin{eqnarray}
u(\phi_\xi,\phi_i)=\sqrt{\alpha^2+2\alpha\eta_i\cos\big(\phi_\xi-\phi_i\big)+\eta_i^2}, \qquad {\rm where}\qquad \alpha=k\sqrt\frac{2r_g}{r}, \qquad \eta_i=k\frac{\rho_i}{f},
  \label{eq:eps}
\end{eqnarray}
where $\alpha$ and $\eta_i$ are the spatial frequencies set by the lens's monopole and that by the imaging telescope, correspondingly; and $(\rho_i,\phi_i)$ are the coordinates of the image sensor, while $f$ is the focal length of the telescope.

Expressions (\ref{eq:psf=})  and (\ref{eq:Pv})  are the PSF$(\vec x)$ of the extended lens and the intensity of light, $I({\vec x},{\vec x}_i)$, observed at the image sensor of an imaging telescope. The optical properties are guided by  (\ref{eq:B2-STF}) and (\ref{eq:BinscER}), correspondingly.  Based on our prior research \cite{Turyshev-Toth:2021-multipoles}, we know that at each order $\ell$ the PSF will exhibit a unique caustic \cite{Turyshev-Toth:2021-caustics} with the cusps yielding bright images to be observed by the telescope \cite{Turyshev-Toth:2021-imaging}.  This result allows for physically consistent modeling of realistic gravitational lenses. Using the intensity of light observed in the image plane, $I({\vec x},{\vec x}_i)$,  given by (\ref{eq:Pv}) with ${\cal A}({\vec x},{\vec x}_i)$ from  (\ref{eq:BinscER})   we can study imaging with an extended lens associated with a generic gravitational potential.

\subsection{Gravitational phase shift of axisymmetric lenses}

Our analysis below relies on the work that we have done in studying axisymmetric lenses. The gravitational potential of such objects may be expressed via an infinite set of zonal harmonics, $J_\ell, \ell\geq 2$, \cite{Turyshev-Toth:2021-multipoles}, yielding the resulting gravitational phase shift (\ref{eq:eik-ph-shift_tpm}) in the following  form:
{}
\begin{eqnarray}
\xi^{\tt al}_b(\vec b,\vec s) ~~=~~
-kr_g\sum_{\ell=2}^\infty\frac{J_\ell}{\ell}\Big(\frac{R}{b}\Big)^\ell \sin^\ell\beta_s\cos[\ell(\phi_\xi-\phi_s)],
\label{eq:eik-ph-axi*}
\end{eqnarray}
where parametrization of $\vec b$ is from (\ref{eq:note-b}), while $\beta_s$ and $\phi_s$ are angles representing the axis of symmetry, $\vec s$, of the lens:
\begin{eqnarray}
{\vec s}&=&(\sin\beta_s\cos\phi_s,\sin\beta_s\sin\phi_s,\cos\beta_s).
\label{eq:note}
\end{eqnarray}

The expression for the gravitational phase delay given by (\ref{eq:eik-ph-axi*}), and its impact on the optical properties of an axisymmetric lens were studied extensively in \cite{Turyshev-Toth:2021-multipoles,Turyshev-Toth:2021-caustics,Turyshev-Toth:2021-imaging,Turyshev-Toth:2021-quartic}. Specifically, in Ref.~\cite{Turyshev-Toth:2021-caustics}, we established the fact that the properties of the caustics are determined by two parameters,  $\alpha$ from (\ref{eq:eps}) and $\beta^{\tt al}_\ell$ that is given as
{}
\begin{eqnarray}
\beta^{\tt al}_\ell ~~=~~ 2kr_g  \frac{J_\ell}{\ell}\Big(\frac{R_\odot }{\sqrt{2r_g\tilde r}}\Big)^\ell\sin^\ell\beta_s.
\label{eq:zerJ}
\end{eqnarray}

In particular, the amplitude of the $\ell$-th caustic, $\rho_\ell$,  is given as
{}
\begin{eqnarray}
\rho_\ell = \ell^2\frac{\beta^{\tt al}_\ell}{\alpha}= \ell\sqrt{2r_g \tilde r}  J_\ell \Big(\frac{R_\odot }{\sqrt{2r_g \tilde r}}\Big)^\ell\sin^\ell\beta_s.
\label{eq:mag}
\end{eqnarray}

\begin{figure}
\centering
\includegraphics[scale=0.91]{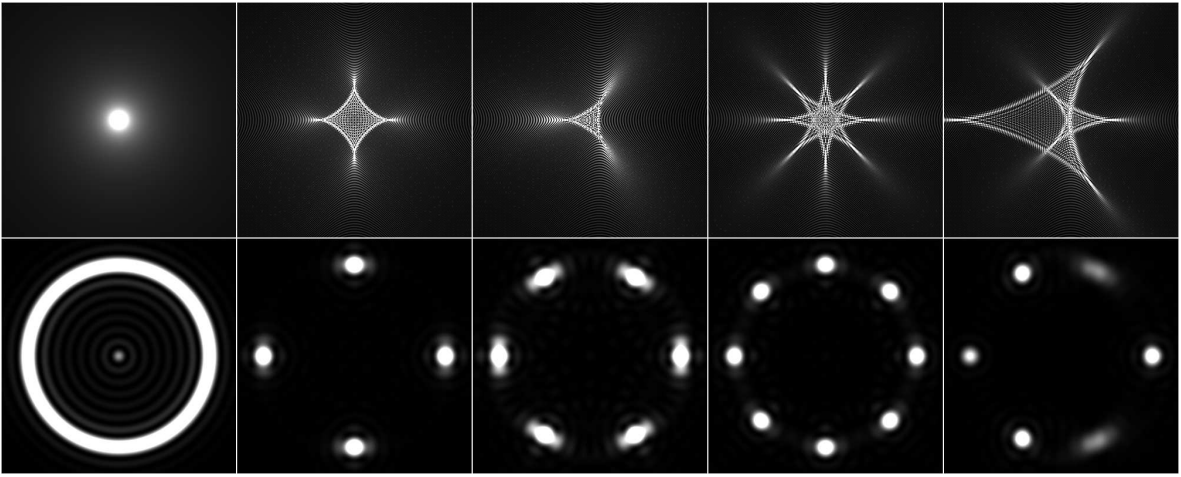}
\caption{\label{fig:montage} Examples of axisymmetric lenses. Top row: Density plot of the PSF of the unperturbed monopole lens; monopole perturbed by quadrupole (characterized by $J_2\ne 0$); monopole perturbed by octupole ($J_3\ne 0$); by hexadecapole ($J_4\ne 0$); and a combination of $J_2$, $J_3$ and $J_4$ perturbations. These density plots represent a cross-section of the light field that is projected by the lens.
Bottom row: The same set of axisymmetric lenses, but with the PSF convolved with that of a thin lens telescope placed at the optical axis, yielding the corresponding telescopic view of light from a point source, as deflected by the lens, including the Einstein-ring for the monopole lens, the Einstein-cross for the quadrupole case and higher-order cases for $J_3$ and $J_4$. The width of the Einstein ring annulus and the size of the light spots produced by the higher-order multipoles are artifacts of the diffraction-limited resolution of the imaging telescope, determined by its aperture and the wavelength of the light. }
\end{figure}

We present this solution for axisymmetric lenses  because the results that we discuss below for generic lenses show the same structure as Eq.~(\ref{eq:eik-ph-axi*}). Thus, the insight that we developed in \cite{Turyshev-Toth:2021-caustics} while studying the optical properties of axisymmetric lenses with  (\ref{eq:eik-ph-axi*}) is directly applicable to the case of compact, extended lenses with arbitrary internal mass distributions. Fig.~\ref{fig:montage} shows the PSF and corresponding telescopic images for several low order zonal harmonics. In particular, we call attention to how the presence of multipole moments breaks up the Einstein-ring into multiple segments. When the impact parameter is large, the relative contribution of the multipole moments is suppressed and the Einstein-ring is recovered. In contrast, when the imaging telescope moves away from the optical axis, the segments are displaced; as the telescope moves outside the caustic region of the PSF, the segments merge into the primary and secondary images that characterize the monopole lens when viewed from an off-axis telescope location.

\subsection{Gravitational phase shift of generic lenses}
\label{sec:phase shift}

Using the intensity of light observed in the image plane, $I({\vec x},{\vec x}_i)$,  given by (\ref{eq:Pv}) with ${\cal A}({\vec x},{\vec x}_i)$ from  (\ref{eq:BinscER}) we can study imaging with an extended gravitational lens that has a generic internal structure and mass distribution. As we see, both expressions, PSF$(\vec x)$ and $I({\vec x},{\vec x}_i)$, depend on the gravitational phase shift (\ref{eq:eik-ph-shift_tpm}) that is accumulated by the EM wave as it travels in the vicinity of a compact massive body. In fact, the properties of a particular mass distribution within that lens are encoded in this gravitational phase shift. Thus, in order for us to evaluate the possibility of extracting the information on the properties of a particular lens, we need to examine images formed by various lenses.

For that, we use (\ref{eq:eik-ph-shift_tpm}), the gravitational eikonal phase shift expressed via the STF multipole moments that, to ${\cal O}(r_g^2)$, was obtained in \cite{Turyshev-Toth:2021-STF-moments}. Our quantity of interest, the eikonal gravitational phase shift $2\xi_b(\vec b,\vec s)$, is obtained by dropping the monopole term from (\ref{eq:eik-ph-shift_tpm}),  or by presenting it as $\varphi(\vec b) =kr_g\ln 4k^2rr_0-2kr_g\ln kb+2\xi_b(\vec b,\vec s)$, which yields the following result:
 {}
\begin{eqnarray}
\xi_b(\vec b)&=& -kr_g\sum_{\ell=2}^\infty
\frac{(-1)^\ell}{\ell!} {\cal T}^{<a_1...a_\ell>}
\hat \partial_{<a_1}... \hat\partial_{a_\ell>}\ln kb \equiv
kr_g\sum_{\ell=2}^\infty
\frac{(2\ell-2)!!}{\ell! \,b^\ell} \sqrt{t^{+2}_\ell +t^{\times2}_\ell}\cos[\ell(\phi_\xi-\phi_\ell)],
\label{eq:eik-ph2-TT}
\end{eqnarray}
where $\hat\partial_a$ in the first from of this expression is the derivative with respect to the vector of the impact parameter, namely $\hat\partial_a=\partial/\partial b^a$, see \cite{Turyshev-Toth:2021-STF-moments} for the details of how to compute  derivatives  $\hat \partial_{<a_1}... \hat\partial_{a_\ell>}\ln kb$ that at each order $\ell$ form a TT-projection operator on the plane orthogonal to $\vec k$.

Using the second expression from (\ref{eq:eik-ph2-TT}), we see that at each order $\ell$, we have a caustic similar to that resulting from (\ref{eq:eik-ph-axi*}), but the amplitude of the $\ell$-th caustic in the general case, $\rho_\ell$,  is given as
{}
\begin{eqnarray}
\beta_\ell=2kr_g
\frac{(2\ell-2)!!}{\ell! \,(\sqrt{2r_g\tilde r})^\ell} \sqrt{t^{+2}_\ell +t^{\times2}_\ell}
\qquad \Rightarrow \qquad
\rho_\ell = \ell^2\frac{\beta_\ell}{\alpha}=
\frac{\ell^2(2\ell-2)!!}{\ell! \,(\sqrt{2r_g\tilde r})^{\ell-1}} \sqrt{t^{+2}_\ell +t^{\times2}_\ell}.
\label{eq:mag-gen}
\end{eqnarray}
Thus,  at each STF order $\ell$, the gravitational phase shift (\ref{eq:eik-ph2-TT}) yields familiar caustics but in this case the caustics are both scaled and rotated. This is in agreement with the fact that there are only two remaining degrees of freedom that are available as a result of the TT-projection on the lens plane. Clearly, applying result (\ref{eq:mag-gen}) to the case of axisymmetric lenses with axis of rotation from (\ref{eq:note}), we recover the result (\ref{eq:mag}).

Although both expressions in (\ref{eq:eik-ph2-TT}) are equivalent, technically it is more straightforward to work with the first one.\footnote{This is because of the fact that we first compute the moments of the STF mass moments ${\cal T}^{<a_1...a_\ell>}$ in a body-fixed coordinates and then rotate the tensor moments to inertial coordinates (see Sec.~\ref{sec:rotate} for details). The tensor notation in the first form of the expressions (\ref{eq:eik-ph2-TT}) (and its expanded form (\ref{eq:eik-ph22p*})--(\ref{eq:eik-ph24*})  for $\ell=2,3,4$) allows us to do that in a more apparent way.}
Furthermore, for our purposes, is sufficient to consider only the lowest order STF mass moments. Accordingly, using the first of the two identical expressions in (\ref{eq:eik-ph2-TT}), the gravitational eikonal phase shifts for the quadrupole ($\ell=2$), octupole ($\ell=3$) and hexadecapole ($\ell=4$) STF multipole mass moments take the form (see details in \cite{Turyshev-Toth:2021-STF-moments}):
{}
\begin{eqnarray}
\xi^{[2]}_b(\vec b)
&=& kr_g
\frac{1}{2b^2} {\cal T}^{<ab>}
\Big\{2m^a m^b+k^ak^b-\delta^{ab}\Big\},
\label{eq:eik-ph22p*}\\
\xi^{[3]}_b(\vec b)
&=& kr_g
\frac{1}{6b^3} {\cal T}^{<abc>}
\Big\{8m^a m^b m^c -2m^a (\delta^{bc}-k^bk^c)-2m^b(\delta^{ac}-k^ak^c)- 2m^c(\delta^{ab}-k^ak^b)\Big\},
\label{eq:eik-ph23*}\\
\xi^{[4]}_b(\vec b)&=& kr_g
\frac{1}{4b^4} {\cal T}^{<abcd>}
\Big\{8m^a m^b m^c m^d+
{\textstyle\frac{1}{3}} \Big(\delta^{bc}-k^bk^c\Big)\Big(\delta^{ad}-k^ak^d\Big)+
{\textstyle\frac{1}{3}} \Big(\delta^{ac}-k^ak^c\Big)\Big(\delta^{bd}-k^bk^d\Big)+
\nonumber\\
&&\hskip 20pt +\,
{\textstyle\frac{1}{3}} \Big(\delta^{ab}-k^ak^b\Big)\Big(\delta^{cd}-k^ck^d\Big)-
{\textstyle\frac{4}{3}}  \Big(m^am^b\big(\delta^{cd}-k^ck^d\big)+
m^am^c\big(\delta^{bd}-k^bk^d\big)+
m^am^d\big(\delta^{bc}-k^bk^c\big)+
\nonumber\\
&&\hskip 20pt +\,
m^bm^c\big(\delta^{ad}-k^ak^d\big)+m^b m^d\big(\delta^{ac}-k^ak^c\big)+m^c m^d\big(\delta^{ab}-k^ak^b\big)\Big)\Big\},
\label{eq:eik-ph24*}
\end{eqnarray}
where $\vec k$ and $\vec m$ are from (\ref{eq:note-k}) and (\ref{eq:note-b}), correspondingly. According to \cite{Turyshev-Toth:2021-STF-moments}, we recognize that the expressions in curly brackets in (\ref{eq:eik-ph22p*})--(\ref{eq:eik-ph24*}) are the TT-projection operators on the plane perpendicular to $\vec k$. Following the approach presented in \cite{Turyshev-Toth:2021-STF-moments} one can easily present these equations in the form  of the second form of the expression in (\ref{eq:eik-ph2-TT}).

Below we will explore the fact established in \cite{Turyshev-Toth:2021-STF-moments} stating that any extended mass distribution will result in gravitational shift that has the structure which is nearly identical to (\ref{eq:eik-ph-axi*}). This is the result that all the STF multiple moments are TT-projected on the lens plane. The procedure allows only for two degrees of freedom.

\subsection{Computing the lowest STF moments}
\label{sec:rotate}

To implement our objectives, we first introduce the STF moments in a particular body-centric coordinate system that is convenient for calculations. For that  we reserve a special notation ${\cal T}_0^L$, while using the usual definition (\ref{eq:mom}):
{}
\begin{eqnarray}
{\cal T}_0^{L}=\int d^3{\vec x} \, \rho({\vec x})\, x^L, \qquad {\rm where} \qquad L\in [1,\ell].
\label{eq:Iab}
\end{eqnarray}
As we shall see below, this definition for STF moments ${\cal T}_0^L$ in a technically convenient coordinate system is related to that ${\cal T}^L$ from (\ref{eq:mom}) in an arbitrary coordinates by a simple rotation.

The coordinate combinations needed to compute the lowest Cartesian STF multipole moments \cite{Soffel-Han:2019} are given as:
{}
\begin{eqnarray}
r^2\hat n_{ij}&=&{\rm STF}_{ij}\Big( x^i x^j\Big)=
x^i x^j-  {\textstyle\frac{1}{3}} r^2\delta^{ij},
\label{eq:sft2}\\
r^3\hat n_{ijk}&=&{\rm STF}_{ijk}\Big( x^i x^j x^k\Big)=
x^i x^j x^k- {\textstyle\frac{1}{5}} r^2\Big(\delta^{ij}x^k+\delta^{jk}x^i+\delta^{ki}x^j\Big),
\label{eq:sft3}\\
r^4\hat n_{ijkl}&=&{\rm STF}_{ijkl}\Big( x^i x^jx^kx^l\Big)=
x^ix^jx^kx^l-\nonumber\\
&-&
{\textstyle\frac{1}{7}} r^2
\Big(x^ix^j\delta^{kl}+x^ix^k\delta^{jl}+x^ix^l\delta^{jk}+x^jx^k\delta^{il}+x^jx^l\delta^{ik}+x^kx^l\delta^{ij}\Big)+
{\textstyle\frac{1}{35}} r^4
\Big(\delta^{ij}\delta^{kl}+\delta^{ik}\delta^{jl}+\delta^{il}\delta^{jk}\Big).~~~
\label{eq:sft4}
\end{eqnarray}

To consider lensing by bodies of known shapes, we use their STF multipole moments. Note that, technically, these moments are easier to compute in their center-of-gravity coordinate system. Thus, we will distinguish two sets of moments: those computed in a particular coordinate system that simplifies the calculations, ${\cal T}_0^L$, introduced in (\ref{eq:Iab}), and those rotated to the arbitrary frame ${\cal T}^L$, used in (\ref{eq:mom}). Physically, these two are identical. Rotating these moments to the chosen coordinate system generally involves the three Euler angles.

We start by taking the reference orientation so that the principal axes coincide with the basis vectors $(e_j)_{j=1,2,3}$. The Euler angles are based on the fact that any general rotation $\bf R$ can be written in terms of three angles and so that $\bf R$ is the composition of three rotations:
\begin{eqnarray}
{\bf R}(\phi_s,\beta_s,\psi)={\bf R}_3(\psi){\bf R}_1(\beta_s){\bf R}_3(\phi_s)=
\begin{pmatrix}
\cos\psi& \sin\psi& 0\\
-\sin\psi& \cos\psi& 0 \\
0 & 0& 1
\end{pmatrix}
\begin{pmatrix}
1& 0& 0\\
0& \cos\beta_s& \sin\beta_s \\
0 & -\sin\beta_s& \cos\beta_s
\end{pmatrix}
\begin{pmatrix}
\cos\phi_s& \sin\phi_s& 0\\
-\sin\phi_s& \cos\phi_s& 0 \\
0 & 0& 1
\end{pmatrix}\equiv R^{ij},
  \label{eq:rot}
\end{eqnarray}
where ${\bf R}_3(\psi)$ is a right-handed rotation of angle $\psi$ around the $x^3$ axis, ${\bf R}_1(\beta_s)$  is a right-handed rotation of angle $\beta_s$ about the $x^1$ axis, and ${\bf R}_3(\phi_s)$ a right-handed rotation of angle $\phi_s$ about the $x^3$ axis. (Note that conventionally, the set of $(\phi,\theta,\psi)$ angles are used to denote the Euler angles. We choose to denote these angles as $(\phi_s,\beta_s,\psi)$ for  consistency with our prior research, e.g.,
\cite{Turyshev-Toth:2021-multipoles,Turyshev-Toth:2021-caustics,Turyshev-Toth:2021-imaging,Turyshev-Toth:2021-all-regions,Turyshev-Toth:2021-quartic,Turyshev-Toth:2021-STF-moments}.)

To rotate the STF multipole moments, ${\cal T}_0^L$, from the body coordinate frame to the chosen coordinate frame and to obtain ${\cal T}^L$, we must rotate these tensors:
\begin{eqnarray}
 {\cal T}^{ij}={R}^i_p R^j_q {\cal T}_0^{pq},
 \qquad
 {\cal T}^{ijk}={R}^i_p R^j_qR^k_s {\cal T}_0^{pqs}
 \qquad
 {\cal T}^{ijkl}={R}^i_p R^j_q R^k_s R^l_w {\cal T}_0^{pqsw},
  \label{eq:rot3}
\end{eqnarray}
where $ {\cal T}^{ij}$, $ {\cal T}^{ijk}$ and $ {\cal T}^{ijkl}$ are the lowest STF multipole tensors transformed to arbitrary inertial coordinates and ${R}^i_j \equiv {\bf R}^T$ being a transpose of ${\bf R}$ from (\ref{eq:rot}). (As such a rotation from body-fixed  to an inertial orientation involves all the components of the tensor $ {\cal T}^{ij}$, using the first expression in (\ref{eq:eik-ph2-TT}) which is written in a tensor form, technically is more convenient, which explains the  choice of (\ref{eq:eik-ph22p*})--(\ref{eq:eik-ph24*})).

\begin{figure}
\vskip 5pt
 \begin{center}
\begin{minipage}[b]{0.35\linewidth}
\centering
 \includegraphics[width=0.50\linewidth]{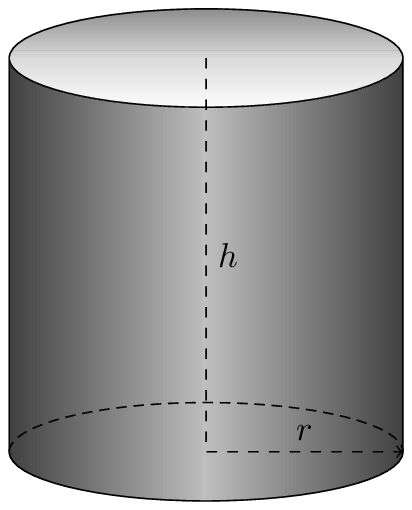}
 \caption{\label{fig:cylinder}The cylinder, parameterized by its radius ($r$) and height ($h$).}
\end{minipage}
  \hskip 10pt
\begin{minipage}[b]{0.35\linewidth}
\centering
 \includegraphics[width=0.53\linewidth]{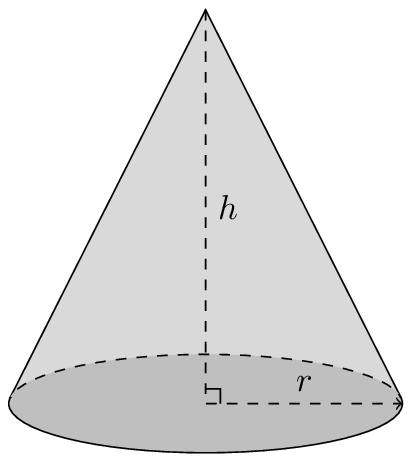}
\caption{\label{fig:rightcone}The right circular cone and its parametrization.}
\end{minipage}
\vskip 16pt
\begin{minipage}[b]{0.35\linewidth}
\vskip -16pt
\centering
 \includegraphics[width=0.80\linewidth]{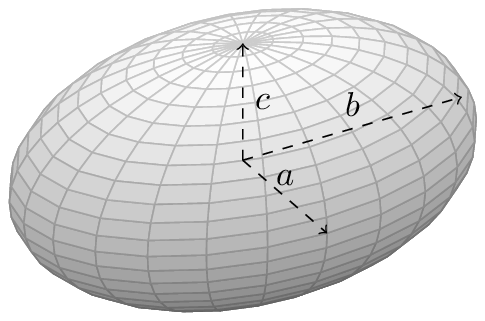}
\caption{\label{fig:ellipsoid}The generic ellipsoid and its parameterization.}
\end{minipage}
  \hskip 10pt
\begin{minipage}[b]{0.35\linewidth}
\centering
 \includegraphics[width=0.60\linewidth]{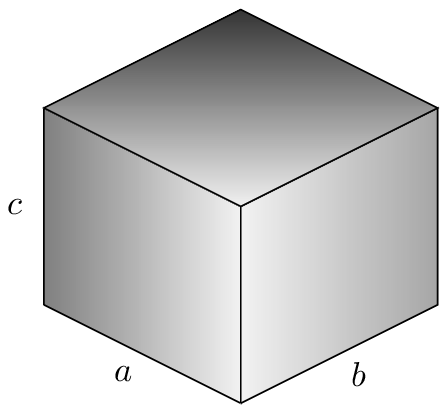}
\caption{\label{fig:cuboid}The generic cuboid characterized by its width ($a$), depth ($b$) and height ($c$).}
\end{minipage}
\vskip 16pt
\begin{minipage}[b]{0.35\linewidth}
\centering
 \includegraphics[width=0.55\linewidth]{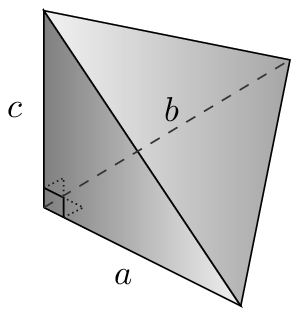}
\caption{\label{fig:tetra}The trirectangular tetrahedron.}
\end{minipage}
 \end{center}
\end{figure}

\section{Lensing with bodies of various shapes}
\label{sec:lens-Plat}

To demonstrate the practical utility of our results, we consider  gravitational lensing by some classic geometric objects with uniform mass density, such as the ellipsoid, the cuboid, the cylinder and the right circular cone, shown in Figs.~\ref{fig:ellipsoid}--\ref{fig:tetra}, correspondingly. Clearly, these are not the lenses that we find in any astronomical observations. However, these exotic body shapes allow us to demonstrate that even these bodies may not be unambiguously distinguished from each other at the level of a particular STF moment. Ultimately, our results suggest that there is missing information about the body mass distribution that is not revealed via gravitational lensing  \cite{Turyshev-Toth:2021-STF-moments}. This is due to the fact that the longitudinal STF mass multipole moments of a gravitational field are not observable. The choice of the simple geometric forms shown in Figs.~\ref{fig:ellipsoid}--\ref{fig:tetra} allows us to clearly demonstrate this important point.

This work allows us to devise a strategy that may be employed in the presence of auxiliary information to determine the mass distribution of an extended gravitational lens. For that purpose, our attention is focused on the intensity distribution  $I({\vec x},{\vec x}_i)$ at the focal plane of a telescope  given by (\ref{eq:Pv}). This is the actual observable that is available through an astronomical instrument: the number of arclets formed along the circumference of the Einstein ring, the symmetries of the distribution of these arclets, their image morphology and the relative brightness of the various peaks. We will pay attention to these characteristics as we study the images formed by various lenses.

Below, we will use expressions (\ref{eq:eik-ph22p*})--(\ref{eq:eik-ph24*}) to study lensing by bodies of various exotic shapes. Such an analysis allows us to emphasize the ambiguity in establishing the precise mass distribution of an extended gravitational lens. We invite the reader not to be intimidated by the lengths of some of the expressions that we obtained. Their lengths notwithstanding, they are directly actionable formulas that can almost be described as calculator-friendly. Notably, they represent an accurate wave-theoretical description of the light field that is intercepted by an observing telescope, and as such, they can be used to model directly the light that is deposited on the image sensor of such an instrument. The accuracy and effectiveness of the approach that we demonstrate here through specific examples may offer a new tool when modeling realistic astrophysical lenses of great complexity.

\subsection{Solid and hollow spheres}

First, we consider a solid sphere of radius $R$ and mass $M$. Using the definition (\ref{eq:Iab}) with (\ref{eq:sft2}) we see that the STF quadrupole mass moment tensor of a sphere vanishes, namely ${\cal T}^{ij}=0.$ Similarly, the STF quadrupole moment tensor of a hollow sphere of radius $R$ and mass $M$  yielding ${\cal T}^{ij}=0$, as expected. It is easy to verify that all higher STF mass moments of solid and hollow spheres also vanish, ${\cal T}^L=0, \ell\geq 1$. Thus, the gravitational potential (\ref{eq:pot_w_0STF2}) and consequently, the gravitational lensing behavior of these two types of objects---solid and hollow spheres---are identical to those of a monopole or a point mass \cite{Turyshev-Toth:2017}, in accordance with Newton's shell theorem. The optical properties of such monopole gravitational lenses are well established and were extensively discussed in the literature, e.g., \cite{Turyshev-Toth:2017,Turyshev-Toth:2019-extend,Turyshev-Toth:2020-image,Turyshev-Toth:2020-extend}.

\subsection{Solid cylinder}
\label{sec:cyl}

We use the definition for the STF moments (\ref{eq:Iab}), STF coordinate combinations from (\ref{eq:sft2})--(\ref{eq:sft4}) and compute the moments of a solid cylinder with uniform matter distribution with radius $r$, height $h$ (Fig.~\ref{fig:cylinder}).
First, we use a coordinate system positioned at its base and compute the monopole moment of such cylinder which constitutes the cylinder's mass $M$:
{}
\begin{eqnarray}
{\cal T}_0^{0}\equiv M=\int d^3{\vec x} \, \rho({\vec x})=\rho \int_0^h dz \int_0^{r}r'dr' \int_0^{2\pi} d\phi_s'=\rho \, \pi r^2 h
\qquad \Rightarrow \qquad
\rho=\frac{M}{\pi r^2 h}.
\label{eq:I0-cyl}
\end{eqnarray}
With the result for the density $\rho$, we now compute the components of the dipole moment ${\cal T}_0^{i}$:
{}
\begin{eqnarray}
\Big\{
{\cal T}_0^{1}, {\cal T}_0^{2}, {\cal T}_0^{3}\Big\}&=&\frac{M}{ \pi r^2 h}\int_0^h dz \int_0^{r}r'dr' \int_0^{2\pi} d\phi_s' \, \Big\{r'\cos\phi_s',r'\sin\phi_s',z-z_0\Big\}=\Big\{0,0,M \Big({\textstyle\frac{1}{2}}h-z_0\Big)\Big\}.
\label{eq:I1-c1cc}
\end{eqnarray}
Therefore, the center of mass of a cylinder is on the $z$-axis at the position of $z_0={\textstyle\frac{1}{2}}h$. Again, with this choice of $z_0$, all the components of  the dipole moment vanish, ${\cal T}_0^i=0$.
With this result for $z_0={\textstyle\frac{1}{2}}h$, using (\ref{eq:sft2}), we compute the STF quadrupole mass moment of a cylinder with uniform matter distribution in the coordinate system positioned at its center of mass:
{}
\begin{eqnarray}
{\cal T}_0^{ij}=  \frac{1}{12}M\Big(r^2-{\textstyle\frac{1}{3}}h^2\Big)
\begin{bmatrix}
1& 0& 0\\
0& 1& 0 \\
0 & 0& -2
\end{bmatrix}.
\label{sec:cyl2}
\end{eqnarray}
To generalize this expression, we rotate ${\cal T}_0^{ij}$ from (\ref{sec:cyl2}) to an arbitrary coordinate system using the quadrupole transformation rule from (\ref{eq:rot3}),  given as
$ {\cal T}^{ij}={R}^i_p R^j_q {\cal T}_0^{pq},$ and substitute the result it into (\ref{eq:eik-ph22p*}). This allows us to derive an expression for the gravitational phase shift introduced by a quadrupole moment of a uniform massive cylinder:
{}
\begin{eqnarray}
\xi^{[2]}_b(\vec b)&=&kr_g\frac{1}{8 b^2}\Big(r^2-{\textstyle\frac{1}{3}}h^2\Big)\sin^2\beta_s\cos[2(\phi_\xi-\phi_s)].
\label{eq:eik-cyl+}
\end{eqnarray}
It is interesting that when $h=\sqrt{3}r$, the quadrupole contribution of the cylinder vanishes and it behaves like a sphere (neglecting contributions from higher multipole moments).

The STF octupole moment moment of a cylinder is computed using (\ref{eq:sft3}), which reveals the fact that  ${\cal T}_0^{ijk}=0$, thus, $\xi^{[3]}_b(\vec b)=0$.

Finally, we compute the STF hexadecapole mass moment of a cylinder using (\ref{eq:sft4}) in its center-of-mass coordinate system, which results in
{}
\begin{eqnarray}
{\cal T}_0^{ijkl}= \frac{M\Big(r^2(r^2-h^2)+{\textstyle\frac{1}{10}}h^4\Big)}{280}
{\small
\begin{bmatrix}
\begin{pmatrix}
3 & 0 & 0\\
0 &1 & 0\\
0 & 0 & -4
\end{pmatrix} &
\begin{pmatrix}
0 & 1 & 0\\
1 & 0 & 0\\
0 & 0 & 0
\end{pmatrix}
&
\begin{pmatrix}
0 & 0 & -4\\
0 & 0 & 0\\
-4 & 0 & 0
\end{pmatrix}\\
\begin{pmatrix}
0 & 1 & 0\\
1  & 0 & 0\\
0 & 0 & 0
\end{pmatrix} &
\begin{pmatrix}
1 & 0 & 0\\
0 & 3 & 0\\
0 & 0 & -4
\end{pmatrix}&
\begin{pmatrix}
0 & 0 & 0\\
0 & 0 & -4\\\
0 & -4 & 0\
\end{pmatrix} \\
\begin{pmatrix}
0 & 0 & -4\\\
0 & 0 & 0\\\
-4 & 0 & 0\
\end{pmatrix} &
\begin{pmatrix}
0 & 0 & 0\\\
0 & 0 & -4\\\
0 & -4 & 0\
\end{pmatrix} &
\begin{pmatrix}
-4 & 0 & 0\\\
0 & -4 & 0\\\
0 & 0 & 8\
\end{pmatrix}
\end{bmatrix}.
}
\label{eq:Q4cyl}
\end{eqnarray}

To generalize this expression, we rotate the STF hexadecapole tensor (\ref{eq:Q4cyl}) to an arbitrary coordinate system by applying the rule from (\ref{eq:rot3}) given as $ {\cal T}^{ijkl}=R^i_pR^j_qR^k_sR^l_w{\cal T}_0^{pqsw}.$ Substituting the result in (\ref{eq:eik-ph24*}), we obtain the following expression for the gravitational phase shift due to the STF hexadecapole mass moment of the cylinder:
{}
\begin{eqnarray}
\xi^{[4]}_b(\vec b)&=& kr_g
\frac{r^2(r^2-h^2)+{\textstyle\frac{1}{10}}h^4}{32 \,b^4}
\sin^4\beta_s \cos[4 (\phi_\xi -\phi_s )],
\label{eq:eik-phs-4cyl}
\end{eqnarray}
which, due to the axial symmetry of the cylinder, is independent on the angle $\psi$, as expected. Clearly, there are higher non-vanishing STF mass multipoles present, but not only are they small, their contribution, being scaled as $1/b^\ell$,  is much suppressed.

\subsection{Right circular cone}
\label{sec:cone}

We consider a right circular cone with radius $r$, height $h$ (Fig.~\ref{fig:rightcone}), and mass $M$ and compute its STF moments in the coordinate system at its center of gravity. Again, we use the definition for the STF moments (\ref{eq:Iab}), the STF coordinate combinations from (\ref{eq:sft2})--(\ref{eq:sft4}), and compute these moments for a cone with uniform density using a coordinate system positioned at its base. First, we compute the monopole moment for the mass of the cone:
{}
\begin{eqnarray}
{\cal T}_0^{0}\equiv M=\int d^3{\vec x} \, \rho({\vec x})=\rho \int_0^h dz \int_0^{(r/h)z}r'dr' \int_0^{2\pi} d\phi_s'=\rho {\textstyle\frac{1}{3}} \pi r^2 h
\qquad \Rightarrow \qquad
\rho=\frac{M}{{\textstyle\frac{1}{3}} \pi r^2 h}.
\label{eq:I0-c}
\end{eqnarray}
With this result, for the density of the mass distribution with  the cone, we compute the dipole moment:
{}
\begin{eqnarray}
\Big\{{\cal T}_0^{1},
{\cal T}_0^{2}, {\cal T}_0^{3}\Big\}&=&\frac{M}{{\textstyle\frac{1}{3}} \pi r^2 h}\int_0^h dz \int_0^{(r/h)z}r'dr' \int_0^{2\pi} d\phi_s' \, \Big\{r' \cos\phi_s', r' \sin\phi_s',z-z_0\Big\}=\Big\{0,0,M\Big({\textstyle\frac{3}{4}}h-z_0\Big)\Big\}.
\label{eq:I1-c1}
\end{eqnarray}
Therefore, the center of mass of a cone is on the $z$-axis at the position of $z_0={\textstyle\frac{3}{4}}h$. With this choice of $z_0$, all the components of  the dipole moment vanish, ${\cal T}_0^i=0$. Next, using this result for $z_0$ and expression (\ref{eq:sft2}), we compute the STF quadrupole mass moment of a right circular cone in the coordinate system at its center of mass:
{}
\begin{eqnarray}
{\cal T}_0^{ij}=  \int d^3{\vec x} \, \rho({\vec x}){\rm STF}_{ij}\big( x^i x^j\big)= \frac{1}{20}M\Big(r^2-{\textstyle\frac{1}{4}}h^2\Big)
\begin{bmatrix}
1& 0& 0\\
0& 1& 0 \\
0 & 0& -2
\end{bmatrix}.
\label{sec:cone2}
\end{eqnarray}

To generalize this expression, we rotate ${\cal T}_0^{ij}$ from (\ref{sec:cone2}) to an arbitrary coordinate system by using the rule $ {\cal T}^{ij}={R}^i_p R^j_q {\cal T}_0^{pq} $
from  (\ref{eq:rot3}) and substitute the result it into (\ref{eq:eik-ph22p*}). This allows us to derive  expression for the gravitational phase shift introduced by gravitational lensing on a quadrupole moment of a regular uniform massive cone:
{}
\begin{eqnarray}
\xi^{[2]}_b(\vec b)&=&kr_g\frac{3}{40}\frac{1}{b^2}\Big(r^2-{\textstyle\frac{1}{4}}h^2\Big)\sin^2\beta_s\cos[2(\phi_\xi-\phi_s)].
\label{eq:eik-cone+}
\end{eqnarray}
Note that this result is similar to that of an axisymmetric ellipsoid (\ref{eq:eik-ax-ellip2}) or a cylinder (\ref{eq:eik-cyl+}).
Also, if $h=2r$, the quadrupole moment (\ref{sec:cone2}) vanishes along with the corresponding eikonal phase shift (\ref{eq:eik-cone+}).

We then compute the STF octupole moment of a right circular cone using (\ref{eq:sft3}), which yields the following result:
{}
\begin{eqnarray}
{\cal T}_0^{ijk}=  \int d^3{\vec x} \, \rho({\vec x}){\rm STF}_{ijk}\big( x^i x^j x^k\big)=\frac{3}{400}M\Big(r^2+{\textstyle\frac{1}{6}}h^2\Big)h
{\small
\begin{bmatrix}
\begin{pmatrix}
0\\
0\\
1
\end{pmatrix} &
\begin{pmatrix}
0\\
0\\
0
\end{pmatrix}
&
\begin{pmatrix}
1\\
0\\
0
\end{pmatrix}\\
\begin{pmatrix}
0\\
0\\
0
\end{pmatrix} &
\begin{pmatrix}
0\\
0\\
1
\end{pmatrix}&
\begin{pmatrix}
0\\
1\\
0
\end{pmatrix} \\
\begin{pmatrix}
1\\
0\\
0
\end{pmatrix} &
\begin{pmatrix}
0\\
1\\
0
\end{pmatrix} &
\begin{pmatrix}
0\\
0\\
-2
\end{pmatrix}
\end{bmatrix}.
}
\label{eq:I3-Qc}
\end{eqnarray}

Again, to generalize the results, we rotate the STF octupole tensor (\ref{eq:I3-Qc}) to an arbitrary coordinate system by applying transformation rule from (\ref{eq:rot3}):
${\cal T}^{ijk}=R^i_pR^j_qR^k_s{\cal T}_0^{pqs}.$ After that, we substitute the result in (\ref{eq:eik-ph23*}) and obtain the following expression for the gravitational phase shift introduced by the octupole of a right circular cone:
{}
\begin{eqnarray}
\xi^{[3]}_b(\vec b)&=& -kr_g
\frac{\big(r^2+{\textstyle\frac{1}{6}}h^2\big)h}{80b^3}
\sin^3\beta_s \sin[3 (\phi_\xi -\phi_s )],
\label{eq:eik-phs-cone}
\end{eqnarray}
which, due to the axial symmetry, is independent on the angle $\psi$, as expected.

Although the right circular cone is axisymmetric, it has no reflection symmetry with respect to the plane of its rotational symmetry (i.e., no ``north--south'' symmetry), which makes this shape particularly interesting as it results in the presence of odd harmonics. When we look at the octupole moment of this object (\ref{eq:I3-Qc}) in the STF representation (\ref{eq:eik-phs-cone}), which we repeat here for convenience, the gravitational phase shift takes the form (\ref{eq:eik-phs-cone}). We indeed find that $\xi^{[2]}$ vanishes when $h=2r$, whereas $\xi^{[3]}$ vanishes only for $h=0$.

Finally, we compute the STF hexadecapole mass moment of a right circular cone using (\ref{eq:sft4}), which yields:
{}
\begin{eqnarray}
{\cal T}_0^{ijkl} &=& \int d^3{\vec x} \, \rho({\vec x}){\rm STF}_{ijkl}\big( x^i x^j x^l x^k\big)= \nonumber\\[-20pt]
&&\hskip 20pt
=\,\frac{3M\big(160r^4-72 r^2h^2+13h^4\big)}{313600}
{\small
\begin{bmatrix}
\begin{pmatrix}
3 & 0 & 0\\
0 &1 & 0\\
0 & 0 & -4
\end{pmatrix} &
\begin{pmatrix}
0 & 1 & 0\\
1 & 0 & 0\\
0 & 0 & 0
\end{pmatrix}
&
\begin{pmatrix}
0 & 0 & -4\\
0 & 0 & 0\\
-4 & 0 & 0
\end{pmatrix}\\
\begin{pmatrix}
0 & 1 & 0\\
1  & 0 & 0\\
0 & 0 & 0
\end{pmatrix} &
\begin{pmatrix}
1 & 0 & 0\\
0 & 3 & 0\\
0 & 0 & -4
\end{pmatrix}&
\begin{pmatrix}
0 & 0 & 0\\
0 & 0 & -4\\\
0 & -4 & 0\
\end{pmatrix} \\
\begin{pmatrix}
0 & 0 & -4\\
0 & 0 & 0\\
-4 & 0 & 0
\end{pmatrix} &
\begin{pmatrix}
0 & 0 & 0\\
0 & 0 & -4\\
0 & -4 & 0
\end{pmatrix} &
\begin{pmatrix}
-4 & 0 & 0\\
0 & -4 & 0\\
0 & 0 & 8
\end{pmatrix}
\end{bmatrix}.
}
\label{eq:Q4cone}
\end{eqnarray}

Next, we rotate the STF hexadecapole tensor (\ref{eq:Q4cone}) to an arbitrary coordinate system by relying on (\ref{eq:rot3}):
${\cal T}^{ijkl}=R^i_pR^j_qR^k_sR^l_w{\cal T}_0^{pqsw}.$
Then, by substituting the result in (\ref{eq:eik-ph24*}), we obtain the following expression for the gravitational phase shift introduced by the STF hexadecapole mass moment of a right circular cone:
{}
\begin{eqnarray}
\xi^{[4]}_b(\vec b)&=& kr_g
\frac{3\big(160r^4-72 r^2h^2+13h^4\big)}{35840 \,b^4}
\sin^4\beta_s \cos[4 (\phi_\xi -\phi_s )],
\label{eq:eik-phs-4cone}
\end{eqnarray}
which, again, due to the axial symmetry of the cone is independent on the angle $\psi$, as expected.

Higher-order moments, which are not calculated here, also contribute of course, but their contribution vanishes rapidly with increasing values of the impact parameter. Therefore, the approximations presented here, in particular the visualizations remain valid so long as $b\gtrsim (r,h)$.

One immediate conclusion from this analysis is that, if we consider only the quadrupole moment, an ambiguity is present in specifying the lens's shape. Inclusion of the next order moment reduces that ambiguity but does not completely eliminate it. Furthermore, we would need to include several moments and operate in the strong lensing regime (in the vicinity of the beginning of the focal region along the optical axis, where the impact parameter is the smallest) to have good constraints on the shape and mass distribution of the lens. We will further discuss this point below when considering bodies of other shapes.

\subsection{Solid ellipsoid}
\label{sec:ellipsoid}
\label{app:stf-ellips}

We now consider an ellipsoid with uniform density distribution. Using a Cartesian coordinate system in which the origin is the center of the ellipsoid and the coordinate axes are its axes, the implicit equation of the ellipsoid has the standard form,
{}
\begin{eqnarray}
\Big(\frac{x}{a}\Big)^2 +\Big(\frac{y}{b}\Big)^2+\Big(\frac{z}{c}\Big)^2 = 1,
\label{eq:ell}
\end{eqnarray}
where $a, b, c $ are positive real numbers.

Next, we use the definition for the STF moments (\ref{eq:Iab}) and expressions (\ref{eq:sft2})--(\ref{eq:sft4}), to compute the STF mass moments of an ellipsoid with a uniform density using coordinate system positioned at its center of mass:
{}
\begin{eqnarray}
{\cal T}_0^{0}=M=\int d^3{\vec x} \, \rho({\vec x})=\rho \int_{-a}^a
dx \int_{-b\sqrt{1-(x/a)^2}}^{b\sqrt{1-(x/a)^2}}dy\int_{-c\sqrt{1-(x/a)^2-(y/b)^2}}^{c\sqrt{1-(x/a)^2-(y/b)^2}}dz.
\label{eq:I0*}
\end{eqnarray}
To compute this triple integral, we change variables as $\{x,y,z\}=\{ax',by',cz'\}$ and have the following result:
{}
\begin{eqnarray}
{\cal T}_0^{0}=M=\int d^3{\vec x} \, \rho({\vec x})=\rho \,abc \int_{-1}^1
dx' \int_{-\sqrt{1-x'^2}}^{\sqrt{1-x'^2}}dy'\int_{-\sqrt{1-x'^2-y'^2}}^{\sqrt{1-x'^2-y'^2}}dz'=\frac{4\pi}{3} \rho \,abc \qquad \Rightarrow \qquad
\rho=\frac{3}{4\pi}\frac{M}{abc}.
\label{eq:I20*}
\end{eqnarray}

Computing the dipole moment is straightforward:
{}
\begin{eqnarray}
{\cal T}_0^{i}=\int d^3{\vec x} \, \rho({\vec x})x^i=\frac{3M}{4\pi} \int_{-1}^1
dx' \int_{-\sqrt{1-x'^2}}^{\sqrt{1-x'^2}}dy'\int_{-\sqrt{1-x'^2-y'^2}}^{\sqrt{1-x'^2-y'^2}}dz'\Big\{ax',by',cz'\Big\}=0.
\label{eq:I1*}
\end{eqnarray}
We see that the dipole moment vanishes in the center-of-mass reference frame, ${\cal T}_0^i=0$.

To compute the  quadrupole, we use the relevant STF expression for coordinate combination given by (\ref{eq:sft2}). As a result, with this, the STF quadrupole moment tensor of a solid ellipsoid with a uniform density distribution, semi-axes $a, b, c$ (Fig.~\ref{fig:ellipsoid}) and mass $M$ is computed to have the following form:
{}
\begin{eqnarray}
{\cal T}_0^{ij} = \int d^3{\vec x} \, \rho({\vec x}){\rm STF}_{ij}\big( x^i x^j\big)=\frac{M}{15}
\begin{bmatrix}
2a^2-b^2-c^2& 0& 0\\
0& 2b^2-a^2-c^2 & 0 \\
0 & 0& 2c^2-a^2-b^2
\end{bmatrix}.
\label{eq:ellipse0}
\end{eqnarray}

Note that this expression is given in a specific coordinate frame with primary components of the moment of inertia. To generalize result (\ref{eq:ellipse0}) and to develop an expression for the gravitational phase shift due to an ellipsoid, we first rotate ${\cal T}_0^{ij}$ from (\ref{eq:ellipse0}) to assume a generic orientation with respect to the incident direction of the EM wave propagation, given by $\vec k$. After that, we rotate ${\cal T}_0^{ij} $ using (\ref{eq:rot3}) and substitute the result in (\ref{eq:eik-ph22p*}), while using the parametrization for $\vec b$ and $\vec k$ from (\ref{eq:note-k})--(\ref{eq:note-x}). To conduct this transformation, we  study rotations of STF tensors and derive expressions to describe arbitrary orientations of a body with respect to the chosen coordinate system. As a result, the gravitational phase shift due to an ellipsoid has the from:
{}
\begin{eqnarray}
\xi^{[2]}_b(\vec b)&=&kr_g\frac{1}{10b^2_0}\Big\{\cos[2(\phi_\xi-\phi_s)]\Big[\Big({\textstyle\frac{1}{2}}\big(a^2+b^2\big)-c^2\Big)\sin^2\beta_s+
 {\textstyle\frac{1}{2}}\big(a^2-b^2\big)
 (1+\cos^2\beta_s)\cos2\psi\Big]+\nonumber\\
&&\hskip 34pt +\,\sin[2(\phi_\xi-\phi_s)]\big(a^2-b^2\big)\sin2\psi\cos\beta_s\Big\},
\label{eq:eik-ellip0}
\end{eqnarray}
where we already see the familiar harmonic structure of the astroid caustic. Specifically, this result  exhibits  the form that we seen in the case of axially-symmetric mass distributions, e.g., \cite{Turyshev-Toth:2021-multipoles,Turyshev-Toth:2021-caustics}, namely
\begin{eqnarray}
\xi^{[2]}_b(\vec b)=kr_g\frac{Q_{\tt e2}}{10b^2_0}\cos[2(\phi_\xi-\phi_s-\phi_{\tt e2})],
\label{eq:eik-ellip}
\end{eqnarray}
where we introduced the magnitude, $Q_{\tt e2}$, and phase, $\phi_{\tt e2}$,  for a generic ellipsoid:
{}
\begin{eqnarray}
Q_{\tt e2}&=&
\Big\{\Big[\sin^2\beta_s\, \Big({\textstyle\frac{1}{2}}\big(a^2+b^2\big)-c^2\Big)+
(1+\cos^2\beta_s)\cos2\psi\, {\textstyle\frac{1}{2}}\big(a^2-b^2\big)\Big]^2+
\Big[\cos\beta_s\sin2\psi\Big(a^2-b^2\Big)\Big]^2\Big\}^\frac{1}{2},
\label{eq:eik-defcc2-a1}\\
\cos2\phi_{\tt e2}&=&\frac{\sin^2\beta_s\, \Big({\textstyle\frac{1}{2}}\big(a^2+b^2\big)-c^2\Big)+
(1+\cos^2\beta_s)\cos2\psi\, {\textstyle\frac{1}{2}}\big(a^2-b^2\big)}{Q_{\tt e2}},\qquad \sin2\phi_{\tt e2}=\frac{\cos\beta_s\sin2\psi\Big(a^2-b^2\Big)}{Q_{\tt e2}},~~~~
\label{eq:eik-defcc2-b1}
\end{eqnarray}
with  $\phi_s, \beta_s,\psi$ being the three Euler angles for an arbitrary rotation. Also, note that for the ellipsoid, to avoid conflicting notation with the size of one of the semi-axes, we use $b_0$ to denote the impact parameter in the denominator.

In the case of axial symmetry, when $a=b$, expressions (\ref{eq:eik-defcc2-a1})--(\ref{eq:eik-defcc2-b1}) reduce to $Q_{\tt e2}=(a^2-c^2)\sin^2\beta_s$ and $\phi_{\tt e2}=0$. This can also be confirmed by substituting $a=b$ into expression (\ref{eq:ellipse0}), which then reduces to
\begin{eqnarray}
{\cal T}^{ij}_{0\, \rm axisym} = \frac{(a^2-c^2)}{5}M
\begin{bmatrix}
{\textstyle\frac{1}{3}}& 0& 0\\
0& {\textstyle\frac{1}{3}}& 0 \\
0 & 0& -{\textstyle\frac{2}{3}}
\end{bmatrix}\simeq\frac{2}{5}Ma^2\Big(\frac{a-c}{a}\Big)
\begin{bmatrix}
{\textstyle\frac{1}{3}}& 0& 0\\
0& {\textstyle\frac{1}{3}}& 0 \\
0 & 0& -{\textstyle\frac{2}{3}}
\end{bmatrix},
\label{sec:el-ax}
\end{eqnarray}
which is consistent with the STF moment of a spheroid (ellipsoid of revolution) \cite{Turyshev-Toth:2021-STF-moments}. To demonstrate this, we use (\ref{eq:rot3}) to rotate ${\cal T}_0^{ij} $ from (\ref{sec:el-ax}) to the needed coordinate frame using (\ref{eq:rot3}). Then, we substitute the result into (\ref{eq:eik-ph22p*}) (or, equivalently, in (\ref{eq:eik-ellip})) and derive an expression for the gravitational phase shift introduced by lensing on a spheroid:
{}
\begin{eqnarray}
\xi^{[2]}_b(\vec b)&=&
-kr_gJ_2\frac{a^2}{2b^2}\sin^2\beta_s\cos[2(\phi_\xi-\phi_s)],
\label{eq:eik-ax-ellip2}
\end{eqnarray}
where the normalized dimensionless quadrupole is $J_2=-{\textstyle\frac{2}{5}}(a-c)/a$, as usual and $b$ now is the impact parameter. This result is known from \cite{Turyshev-Toth:2021-multipoles}, were we studied the case of a lens with axial symmetry.

Next, using the combination (\ref{eq:sft3}), we compute the octupole moment of an ellipsoid to see that all components of the STF octupole mass moment vanish ${\cal T}_0^{ijk}=0$ and, thus,  $\xi^{[3]}_b(\vec b)=0$.

Therefore, the next non-vanishing moment is the hexadecapole. With combination (\ref{eq:sft4}), we compute the hexadecapole moment of an ellipsoid in the coordinate system at its center of mass, which results in
{}
\begin{eqnarray}
{\cal T}_0^{ijkl}= \int d^3{\vec x} \, \rho({\vec x}){\rm STF}_{ijkl}\big( x^i x^jx^k x^l\big)=
{\small
\begin{bmatrix}
\begin{pmatrix}
{\tt A} & 0 & 0\\
0 &{\tt AB} & 0\\
0 & 0 & {\tt AC}
\end{pmatrix} &
\begin{pmatrix}
0 & {\tt AB} & 0\\
{\tt AB} & 0 & 0\\
0 & 0 & 0
\end{pmatrix}
&
\begin{pmatrix}
0 & 0 & {\tt AC}\\
0 & 0 & 0\\
{\tt AC} & 0 & 0
\end{pmatrix}\\
\begin{pmatrix}
0 & {\tt AB} & 0\\
{\tt AB}  & 0 & 0\\
0 & 0 & 0
\end{pmatrix} &
\begin{pmatrix}
{\tt AB} & 0 & 0\\
0 & {\tt B} & 0\\
0 & 0 & {\tt BC}
\end{pmatrix}&
\begin{pmatrix}
0 & 0 & 0\\
0 & 0 & {\tt BC}\\\
0 & {\tt BC} & 0\
\end{pmatrix} \\
\begin{pmatrix}
0 & 0 & {\tt AC}\\\
0 & 0 & 0\\\
{\tt AC}& 0 & 0\
\end{pmatrix} &
\begin{pmatrix}
0 & 0 & 0\\\
0 & 0 & {\tt BC}\\\
0 & {\tt BC} & 0\
\end{pmatrix} &
\begin{pmatrix}
{\tt AC} & 0 & 0\\\
0 & {\tt BC} & 0\\\
0 & 0 & {\tt C} \
\end{pmatrix}
\end{bmatrix},
}
\label{eq:Q4ellipse0}
\end{eqnarray}
where the 6 non-vanishing components $\{\tt A, B, C, AB, AC, BC\}$ have the following form:
{}
\begin{eqnarray}
{\tt A}&=&\frac{3M}{1225}\Big(8a^2\big(a^2-b^2-c^2\big)+3b^4+2b^2c^2+3c^4\Big), \qquad
{\tt AB}=-\frac{3M}{1225}\Big(4\big(a^4+b^4\big)-9a^2b^2-c^4+c^2\big(a^2+b^2\big)\Big),\\
{\tt B}&=&\frac{3M}{1225}\Big(8b^2\big(b^2-a^2-c^2\big)+3a^4+2a^2c^2+3c^4\Big), \qquad
{\tt AC}=-\frac{3M}{1225}\Big(4\big(a^4+c^4\big)-9a^2c^2-b^4+b^2\big(a^2+c^2\big)\Big), \\
{\tt C}&=&\frac{3M}{1225}\Big(8c^2\big(c^2-a^2-b^2\big)+3a^4+2a^2b^2+3b^4\Big), \qquad
{\tt BC}=-\frac{3M}{1225}\Big(4\big(b^4+c^4\big)-9b^2c^2-a^4+a^2\big(b^2+c^2\big)\Big).
\label{eq:Q4ellipseABC}
\end{eqnarray}

To generalize the result, we rotate the STF hexadecapole tensor (\ref{eq:Q4ellipse0})--(\ref{eq:Q4ellipseABC}) to an arbitrary coordinate system by applying the transformation rule from (\ref{eq:rot3}), given as  ${\cal T}^{<ijkl>}=R^i_pR^j_qR^k_lR^l_n{\cal T}_0^{<pqln>}.$ Then, we substitute the result in (\ref{eq:eik-ph24*}) and obtain the following expression for the eikonal gravitational phase shift introduced by the hexadecapole moment of an ellipsoid with uniform mass density distribution:
{}
\begin{eqnarray}
\xi^{[4]}_b(\vec b)&=&
\frac{3\,kr_g}{140\,b^4_0}
\Big\{\cos[4 (\phi_\xi -\phi_s)]\Big[{\textstyle\frac{1}{2}}(a^2-b^2)^2\Big(\big(\cos^2\beta_s+{\textstyle\frac{1}{4}}(1+\cos^2\beta_s)^2\big)\cos4\psi+{\textstyle\frac{3}{4}}\sin^4\beta_s\Big)+\nonumber\\
&&\hskip 45pt
+\,\big(a^2-c^2\big)\big(b^2-c^2\big)\sin^4\beta_s+{\textstyle\frac{1}{2}}\big(a^2-b^2\big)\big(a^2+b^2-2c^2\big)\cos2\psi\big(1+\cos^2\beta_s\big)\sin^2\beta_s \Big]+\nonumber\\
&+&
\sin[4 (\phi_\xi -\phi_s)]\big(a^2-b^2\big)\Big(\big(a^2+b^2-2c^2\big)\sin^2\beta_s+\big(a^2-b^2\big)\big(1+\cos^2\beta_s\big)\cos2\psi\Big)\sin2\psi\cos\beta_s \Big\}.
\label{eq:eik-ellipsoid4}
\end{eqnarray}

Therefore, the eikonal gravitational phase shift introduced by the hexadecapole of an ellipsoid with uniform density (\ref{eq:eik-ellipsoid4}) takes the familiar harmonic structure:
\begin{eqnarray}
\xi^{[4]}_b(\vec b)=kr_g \frac{3\,Q_{\tt e4}}{140\,b^4_0} \cos[4(\phi_\xi-\phi_s-\phi_{\tt e4})],
\label{eq:eik-ellip=}
\end{eqnarray}
where the magnitude, $Q_{\tt e4}$, and phase, $\phi_{\tt e4}$,  can readily be read-off directly from (\ref{eq:eik-ellipsoid4}).

\subsection{Solid cuboid}
\label{sec:cuboid}
\label{app:stf-cuboid}

Now we consider the STF moments of a solid homogeneous rectangular block of width $a$, depth $b$, height $c$ (Fig.~\ref{fig:cuboid}).
We use the definition for the STF moments (\ref{eq:Iab}) and expressions (\ref{eq:sft2})--(\ref{eq:sft4}), to compute STF mass moments of a cuboid with a uniform density using a coordinate system positioned at its center of mass:
{}
\begin{eqnarray}
{\cal T}_0^{0}=M=\int d^3{\vec x} \, \rho({\vec x})=\rho \int_{-a/2}^{a/2} dx \int_{-b/2}^{b/2}dy\int_{-c/2}^{c/2}dz = \rho \,abc \qquad \Rightarrow \qquad
\rho=\frac{M}{abc}.
\label{eq:I20cu*}
\end{eqnarray}

Computation of the dipole moment is straightforward and is done in the manner similar to (\ref{eq:I1*}). By doing so, we can easily verify that in the center-of-mass coordinates frame the dipole moment of a cuboid vanishes, ${\cal T}_0^i=0$.

To compute the quadrupole, we again use the corresponding STF expression for the coordinate combination given by (\ref{eq:sft2}). As a result, the STF quadrupole moment of a solid homogeneous rectangular block of width $a$, depth $b$, height $c$ and mass $M$ in a body coordinate frame at is center of mass and oriented along the coordinate axes, has the form
{}
\begin{eqnarray}
{\cal T}_0^{ij} = \int d^3{\vec x} \, \rho({\vec x}){\rm STF}_{ij}\big( x^i x^j\big)=\frac{M}{36}
\begin{bmatrix}
2a^2-b^2-c^2& 0& 0\\
0& 2b^2-a^2-c^2 & 0 \\
0 & 0& 2c^2-a^2-b^2
\end{bmatrix}.
\label{eq:cuboid0}
\end{eqnarray}
It is a bit unexpected but, except for the numerical coefficient, this expression for the STF quadrupole mass moment of a cuboid (\ref{eq:cuboid0}) is identical to that of the ellipsoid given by (\ref{eq:ellipse0}). For a generic cuboid, we obtain an expression for the gravitational eikonal phase shift by substituting the components ${\cal T}_0^{ij}$ from (\ref{eq:cuboid0}) into (\ref{eq:rot3}) and then into (\ref{eq:eik-ph22p*}).
With this result and using (\ref{eq:eik-ph22p*}) and (\ref{eq:rot3}), we have the following result for the gravitational eikonal phase shift
{}
\begin{eqnarray}
\xi^{[2]}_b(\vec b)&=&kr_g\frac{1}{24 b^2_0}\Big\{\cos[2(\phi_\xi-\phi_s)]\Big[\Big({\textstyle\frac{1}{2}}\big(a^2+b^2\big)-c^2\Big)\sin^2\beta_s+
 {\textstyle\frac{1}{2}}\big(a^2-b^2\big)
 (1+\cos^2\beta_s)\cos2\psi\Big]+\nonumber\\
&&\hskip 35pt +\,\sin[2(\phi_\xi-\phi_s)]\big(a^2-b^2\big)\sin2\psi\cos\beta_s\Big\},
\label{eq:eik-cuboid1}
\end{eqnarray}
where we already see the familiar harmonic structure of the astroid caustic. Note that the structure of this expression, except for the amplitude, is identical to that derived for an ellipsoid (\ref{eq:eik-ellip0}). This result may be given again in the familiar form:
\begin{eqnarray}
\xi^{[2]}_b(\vec b)=kr_g\frac{Q_{\tt e2}}{24 b^2_0}\cos[2(\phi_\xi-\phi_s-\phi_{\tt e2})],
\label{eq:eik-cuboid0}
\end{eqnarray}
where the magnitude, $Q_{\tt e2}$, and phase, $\phi_{\tt e2}$, are given by (\ref{eq:eik-defcc2-a1})--(\ref{eq:eik-defcc2-b1}) and we again use $b_0$ to denote the impact parameter, as in (\ref{eq:eik-ellip}). In (\ref{eq:eik-cuboid0}) we recognize the familiar zonal harmonic form of the quadrupole moment, with the axis rotated by $(\phi_s+\phi_{\tt e2})$. Comparing this result (\ref{eq:eik-cuboid0})  to that of an ellipsoid given by (\ref{eq:eik-ellip}), we see that, as expected, the eikonal phase shift induced by a generic cuboid behaves similarly.
Thus, the caustic in the PSF of the gravitating lens of an ellipsoid is a function of two parameters, the magnitude $Q_{e2}$ and rotation angle  $(\phi_s+\phi_{\tt e2})$ that are given by (\ref{eq:eik-defcc2-a1})--(\ref{eq:eik-defcc2-b1}).

Note that the STF mass octupole moment of a cuboid vanishes, ${\cal T}_0^{ijk}=0$, thus $\xi^{[3]}_b(\vec b)=0$.
Therefore, the next non-vanishing term in the external gravitational potential produced by a cuboid will be that due to the hexadecapole.

With the combination (\ref{eq:sft4}), we compute the hexadecapole moment of a cuboid in the coordinate system at its center of mass, which results in a structure identical to that of an ellipsoid, (\ref{eq:Q4ellipse0}), with the 6 non-vanishing components $\{\tt A, B, C, AB, AC, BC\}$ given as:
{}
\begin{eqnarray}
{\tt A}&=&\frac{M}{8400}\Big(24a^4-40a^2\big(b^2+c^2\big)+9b^4+10b^2c^2+9c^4\Big), \qquad
{\tt AB}=-\frac{M}{8400}\Big(12\big(a^4+b^4\big)-45a^2b^2+5c^2\big(a^2+b^2\big)-3c^4\Big),\\
{\tt B}&=&\frac{M}{8400}\Big(24b^4-40b^2\big(a^2+c^2\big)+9a^4+10a^2c^2+9c^4\Big), \qquad
{\tt AC}=-\frac{M}{8400}\Big(12\big(a^4+c^4\big)-45a^2c^2+5b^2\big(a^2+c^2\big)-3b^4\Big),\\
{\tt C}&=&\frac{M}{8400}\Big(24c^4-40c^2\big(a^2+b^2\big)+9a^4+10a^2b^2+9b^4\Big), \qquad
{\tt BC}=-\frac{M}{8400}\Big(12\big(b^4+c^4\big)-45b^2c^2+5a^2\big(b^2+c^2\big)-3a^4\Big).
\label{eq:Q4cuboidABC}
\end{eqnarray}

To generalize this result, we rotate the STF hexadecapole tensor (\ref{eq:Q4ellipse0})--(\ref{eq:Q4ellipseABC}) to an arbitrary coordinate system by applying the transformation rule from (\ref{eq:rot3}), given as
 ${\cal T}^{<ijkl>}=R^i_pR^j_qR^k_lR^l_n{\cal T}_0^{<pqln>}.$ Then, we substitute the result in (\ref{eq:eik-ph24*}) and obtain the following expression for the eikonal gravitational phase shift introduced by the hexadecapole of an ellipsoid:
{}
\begin{eqnarray}
\xi^{[4]}_b(\vec b)&=&
\frac{kr_g}{960\,b^4_0}
\Big\{\cos[4 (\phi_\xi -\phi_s)]\Big[{\textstyle\frac{1}{2}}\big(a^2-b^2\big)\big(3a^2+3b^2-10c^2\big)\big(1+\cos^2\beta_s\big)\sin^2\beta_s\cos2\psi+\nonumber\\
&&\hskip -50pt
+\,{\textstyle\frac{1}{8}}\Big(9\big(a^4+b^4\big)+10a^2b^2+24c^4-40c^2\big(a^2+b^2\big)\Big)\sin^4\beta_s+{\textstyle\frac{1}{8}}\big(3a^4-10a^2b^2+3b^4\big)\Big(4\cos^2\beta_s+\big(1+\cos^2\beta_s\big)^2\Big)\cos4\psi\Big]+\nonumber\\
&&\hskip -50pt +\,
\sin[4 (\phi_\xi -\phi_s)]\Big(\big(a^2-b^2\big)\big(3a^2+3b^2-10c^2\big)\sin^2\beta_s+\big(3a^4-10a^2b^2+3b^4\big)\big(1+\cos^2\beta_s\big)\cos2\psi\Big)\sin2\psi\cos\beta_s\Big\}.
\label{eq:eik-cuboid4}
\end{eqnarray}
Thus, as in the case of an ellipsoid, the eikonal gravitational phase shift introduced by the hexadecapole of a cuboid with uniform density takes the familiar harmonic structure
\begin{eqnarray}
\xi^{[4]}_b(\vec b)=kr_g \frac{Q_{\tt cu4}}{960\,b^4_0} \cos[4(\phi_\xi-\phi_s-\phi_{\tt cu4})],
\label{eq:eik-cuboid}
\end{eqnarray}
where the magnitude, $Q_{\tt cu4}$, and phase, $\phi_{\tt cu4}$,  can be readily read off directly from (\ref{eq:eik-cuboid4}).

We can thus can see that, although the quadrupole moments of the ellipsoid and cuboid introduce very similar structures of the eikonal phase shift, at the level of the hexadecapole their contributions are different.

\subsection{Trirectangular tetrahedron}
\label{sec:tri-tetra}
\label{sec:terta}

The level of degeneracy between shapes such as the cuboid and the ellipsoid, or the cylinder vs. the right circular cone, may perhaps be explained by the fact that all these shapes have either rotational or ``north--south'' symmetry, or both. For this reason, we also opted to investigate a shape that has neither. We picked for this purpose the trirectangular tetrahedron. Despite its lack of basic symmetries, this shape is nonetheless simple enough to be investigated analytically, thus advancing our investigation of observables available in gravitational lensing.

To compute the STF moments of a trirectangular tetrahedron with uniform density, we use the definition for the STF moments (\ref{eq:Iab}) and expressions (\ref{eq:sft2})--(\ref{eq:sft4}). To define the tetrahedron we used the intercept form formula which is $x/a+y/b+z/c=1$, where $a,b,c$ are $x,y,z$ intercepts. Then, the mass is computed to be
{}
\begin{eqnarray}
{\cal T}_0^{0}\equiv M=\int d^3{\vec x} \, \rho({\vec x})=\rho \int_0^a dx \int_0^{b(1-x/a)}dy \int_0^{c(1-x/a-y/b)} dz=\rho {\textstyle\frac{1}{6}} abc
\qquad \Rightarrow \qquad
\rho=\frac{M}{{\textstyle\frac{1}{6}} abc}.
\label{eq:I0-tet}
\end{eqnarray}

With this, we compute the dipole moment:
{}
\begin{eqnarray}
\Big\{{\cal T}_0^{1},{\cal T}_0^{2},{\cal T}_0^{3}\Big\}&=&\frac{M}{{\textstyle\frac{1}{6}} abc} \int_0^a dx \int_0^{b(1-x/a)}dy \int_0^{c(1-x/a-y/b)} dz\Big\{x-x_0,y-y_0,z-z_0\Big\}=
M\Big\{{\textstyle\frac{1}{4}}a-x_0,{\textstyle\frac{1}{4}}b-y_0,{\textstyle\frac{1}{4}}c-z_0\Big\}.
\label{eq:I1-tet1}
\end{eqnarray}
Therefore, the center of gravity of a trirectangular tetrahedron is at the point with coordinates $\vec x_0=\{{\textstyle\frac{1}{4}}a,{\textstyle\frac{1}{4}}b,{\textstyle\frac{1}{4}}c\}$. With this choice of $\vec x_0$, all the components of  the dipole moment vanish, ${\cal T}_0^i=0$.

Using a coordinate system positioned at the center of mass of a solid trirectangular tetrahedron with width $a$, depth $b$, hight $c$ (Fig.~\ref{fig:tetra}), and mass $M$, we compute its STF quadrupole moment in a body coordinate frame at is center of mass and oriented along the coordinate axes:
{}
\begin{eqnarray}
{\cal T}_0^{ij}= \int d^3{\vec x} \, \rho({\vec x}){\rm STF}_{ij}\big( (x^i-x_0^i) (x^j-x_0^j)\big)=\frac{M}{80}
\begin{bmatrix}
2a^2-b^2-c^2& -ab& -ac\\
-ab& 2b^2-a^2-c^2& -bc \\
-ac & -bc& 2c^2-a^2-b^2
\end{bmatrix}.
\label{eq:tetra2}
\end{eqnarray}

For a generic tetrahedron, we obtain an expression for the gravitational eikonal phase shift by substituting the components ${\cal T}_0^{ij}$ from (\ref{eq:tetra2}) into (\ref{eq:rot3}) and then into (\ref{eq:eik-ph22p*}). As a result, we have the following eikonal phase shift:
{}
\begin{eqnarray}
\xi^{[2]}_b(\vec b)&=&kr_g\frac{3}{160 b^2_0}\Big\{\cos[2(\phi_\xi-\phi_s)]\Big[\Big({\textstyle\frac{1}{2}}\big(a^2+b^2\big)-c^2\Big)\sin^2\beta_s+
{\textstyle\frac{1}{2}}\big(a^2-b^2\big)(1+\cos^2\beta_s)\cos2\psi\Big]+
\nonumber\\
&&\hskip 40pt+ \,
\sin[2(\phi_\xi-\phi_s)]\big(a^2-b^2\big)\sin2\psi\cos\beta_s\Big\}+\nonumber\\
&+&kr_g\frac{3}{160 b^2_0}\Big\{\cos[2(\phi_\xi-\phi_s)]\Big[
{\textstyle\frac{1}{3}}ab\big(1+\cos^2\beta_s\big)\sin2\psi-{\textstyle\frac{1}{3}}c\big(a\sin\psi+b\cos\psi\big)\sin2\beta_s\Big]+
\nonumber\\
&&\hskip 40pt+ \,
\sin[2(\phi_\xi-\phi_s)]\Big[{\textstyle\frac{2}{3}}c \big(a\cos\psi-b\sin\psi\big)\sin\beta_s-{\textstyle\frac{2}{3}}ab\cos2\psi\cos\beta_s\Big]\Big\},
\label{eq:eik-tetra2}
\end{eqnarray}
where again we used $b_0$ to denote the impact parameter, which is nearly identical to the ellipsoid with (\ref{eq:eik-defcc2-a1})--(\ref{eq:eik-defcc2-b1}). Note that the structure of the first two lines of this expression is identical to that of (\ref{eq:eik-cuboid1}) and is due to diagonal components of the quadrupole SRF tensor (\ref{eq:tetra2}). The last two lines of this expression are due to non-diagonal components that are present because of the broken north--south symmetry. In any event, expression (\ref{eq:eik-tetra2}) may be cast in the familiar harmonic structure:
\begin{eqnarray}
\xi^{[2]}_b(\vec b)&=&kr_g \frac{3\,Q_{\tt t2}}{160\,b^2_0} \cos[2(\phi_\xi-\phi_s-\phi_{\tt t2})],
\label{eq:eik-tetra2=}
\end{eqnarray}
where the magnitude, $Q_{\tt t4}$, and phase, $\phi_{\tt t4}$,  can readily be read off directly from (\ref{eq:eik-tetra2}).

We have computed the STF octupole mass moments of the trirectangular tetrahedron. Using coordinate system positioned at the center of mass of a solid trirectangular tetrahedron, we compute its STF octupole moment in a body coordinate frame at is center of mass and oriented along the coordinate axes:
{}
\begin{eqnarray}
{\cal T}_0^{ijk}&=& \int d^3{\vec x} \, \rho({\vec x}){\rm STF}_{ij}\big( (x^i-x_0^i) (x^j-x_0^j)(x^k-x_0^k)\big)=\nonumber\\
&=&\frac{M}{2400}
{\small
\begin{bmatrix}
\begin{pmatrix}
3a(2a^2+b^2+c^2)\\
-b(4a^2+3b^2-c^2)\\
-c(4a^2-b^2+3c^2)
\end{pmatrix} &
\begin{pmatrix}
-b(4a^2+3b^2-c^2)\\
-a(3a^2+4b^2-c^2)\\
5abc
\end{pmatrix}
&
\begin{pmatrix}
-c(4a^2-b^2+3c^2)\\
5abc\\
-a(3a^2-b^2+4c^2)
\end{pmatrix}\\
\begin{pmatrix}
-b(4a^2+3b^2-c^2)\\
-a(3a^2+4b^2-c^2)\\
5abc
\end{pmatrix} &
\begin{pmatrix}
-a(3a^2+4b^2-c^2)\\
3b(a^2+2b^2+c^2)\\
c(a^2-4b^2-3c^2)
\end{pmatrix}&
\begin{pmatrix}
5abc\\
c(a^2-4b^2-2c^2)\\
b(a^2-3b^2-4c^2)
\end{pmatrix} \\
\begin{pmatrix}
-c(4a^2-b^2+4c^2)\\
5abc\\
-a(3a^2-b^2+4c^2)
\end{pmatrix} &
\begin{pmatrix}
5abc\\
c(a^2-4b^2-3c^2)\\
b(a^2-3b^2-4c^2)
\end{pmatrix} &
\begin{pmatrix}
-a(3a^2-b^2+4c^2)\\
b(a^2-3b^2-4c^2)\\
3c(a^2+b^2+c^2)
\end{pmatrix}
\end{bmatrix}.
}
\label{eq:tetra3}
\end{eqnarray}

For a generic tetrahedron, we obtain an expression for the gravitational eikonal phase shift by substituting the components ${\cal T}_0^{ijk}$ from (\ref{eq:tetra3}) into (\ref{eq:rot3}) and then into (\ref{eq:eik-ph23*}). Again, to generalize the results, we rotate the STF octupole tensor (\ref{eq:tetra3}) to an arbitrary coordinate system by using the rule from (\ref{eq:rot3}):
${\cal T}^{ijk}=R^i_pR^j_qR^k_s{\cal T}_0^{pqs}.$
After that, we substitute the result in (\ref{eq:eik-ph23*}) and obtain the following expression for the gravitational phase shift introduced by the octupole of a trirectangular tetrahedron:
{}
\begin{eqnarray}
\xi^{[3]}_b(\vec b)&=&\frac{kr_g}{2400\,b^3_0}\Big\{\cos[3(\phi_\xi-\phi_s)]\Big[\Big(a\big(3a^2-b^2+4c^2\big)\cos\psi+b\big(a^2-3b^2-4c^2\big)\sin\psi\Big)\sin^2\beta_s+\nonumber\\
&&+\,5c\Big(ab\cos2\psi-{\textstyle\frac{1}{2}}\big(a^2-b^2\big)\sin2\psi\Big)\sin2\beta_s-
b\big(a^2+2b^2+c^2\big)\big(\sin^2\psi-3\cos^2\beta_s\cos^2\psi\big)\sin\psi+
\nonumber\\
&&\hskip -40pt
+\,a\big(2a^2+b^2+c^2\big)\big(\cos^2\psi-3\cos^2\beta_s\sin^2\psi\big)\cos\psi +a\big(3a^2+4b^2-c^2\big)\Big({\textstyle\frac{1}{4}}\big(\cos\psi+3\cos3\psi\big)\cos^2\beta_s-\cos\psi\sin^2\psi\Big)+\nonumber\\
&&+\,
b\big(4a^2+3b^2-c^2\big)\Big(\cos^2\psi\sin\psi-{\textstyle\frac{1}{4}}\big(\sin\psi-3\sin3\psi\big)\cos^2\beta_s\Big)\Big]+\nonumber\\
&&\hskip -40pt+ \,
\sin[3(\phi_\xi-\phi_s)]\Big[{\textstyle\frac{5}{2}}c\big(a^2+b^2+2c^2\big)\sin^3\beta_s+
5c\Big({\textstyle\frac{1}{2}}c\big(a^2-b^2\big)\cos2\psi+5ab\sin2\psi \Big)\big(1+\cos^2\beta_s\big)\sin\beta_s+\nonumber\\
&&
+\,\Big(b\big(3b^2-a^2+4c^2\big)\cos\psi+a\big(3a^2-b^2+4c^2\big)\sin\psi\Big){\textstyle\frac{1}{2}}\sin2\beta_s\sin\beta_s+\nonumber\\
&&\hskip -40pt
+\,\Big(b\big(a^2+2b^2+c^2\big)\big(3\sin^2\psi-\cos^2\beta_s\cos^2\psi\big)\cos\psi+a\big(2a^2+b^2+c^2\big)\big(3\cos^2\psi-\cos^2\beta_s\sin\psi\big)\sin\psi\Big)\cos\beta_s+\nonumber\\
&&+\,{\textstyle\frac{1}{4}}a\big(3a^2+4b^2-c^2\big)\big(4\cos^2\beta_s\cos^2\psi\sin\psi+3\sin3\psi-\sin\psi\big)\cos\beta_s-\nonumber\\
&&-
\,{\textstyle\frac{1}{4}}b\big(4a^2+3b^2-c^2\big)\big(\cos\psi+3\cos3\psi-4\cos^2\beta_s\cos\psi\sin^2\psi\big)\cos\beta_s\Big]\Big\},
\label{eq:eik-tetra3}
\end{eqnarray}
where the harmonic structure of the result is obvious. In fact,   (\ref{eq:eik-tetra3}) may be cast in the familiar form:
\begin{eqnarray}
\xi^{[3]}_b(\vec b)=kr_g \frac{Q_{\tt t3}}{2400\,b^3_0} \cos[3(\phi_\xi-\phi_s-\phi_{\tt t3})],
\label{eq:eik-tetra3=}
\end{eqnarray}
where the magnitude, $Q_{\tt t3}$, and phase, $\phi_{\tt t3}$,  can readily be read-off directly from (\ref{eq:eik-tetra3}).

We have computed the STF hexadecapole of a tetrahedron to verify that all of its $3\times3\times3\times3=81$ components are non-vanishing, making the results rather lengthy. Because of this, we will not present the hexadecapole moment of the trirectangular tetrahedron here, but show only the gravitational phase shift. Again, to generalize this expression, we rotate the STF hexadecapole tensor  to an arbitrary coordinate system using  the rule ${\cal T}^{ijkl}=R^i_pR^j_qR^k_sR^l_w{\cal T}_0^{pqsw}$
from (\ref{eq:rot3}) and obtain the following expression for the gravitational phase shift introduced by the STF hexadecapole mass moment of a trirectangular tetrahedron:
{}
\begin{eqnarray}
\xi^{[4]}_b(\vec b)&=&\frac{kr_g}{35840 \,b_0^4} \Big\{\cos[4(\phi_\xi-\phi_s)]\Big[{\textstyle\frac{1}{8}}\Big(117\big(a^4+b^4\big)+74a^2b^2+312c^4-296c^2\big(a^2+b^2\big)\Big)\sin^4\beta_s+\nonumber\\
&&\hskip 35pt+ \,
\Big(bc\big(a^2-39b^2+52c^2\big)\cos\psi-ac\big(39a^2-b^2-52c^2\big)\sin\psi\Big)\cos\beta_s\sin^3\beta_s+
\nonumber\\
&&\hskip 0pt+ \,
\Big({\textstyle\frac{1}{2}}\big(a^2-b^2\big)\big(39(a^2+b^2)-74c^2\big)\cos2\psi+ab\big(13(a^2+b^2)+2c^2\big)\sin2\psi\Big)\big(1+\cos^2\beta_s\big)\sin^2\beta_s+\nonumber\\
&&\hskip 0pt+ \,
{\textstyle\frac{1}{8}}\Big(bc\big(a^2+13b^2\big)\cos3\psi-ac\big(13a^2+b^2\big)\sin3\psi\Big)\big(14\sin2\beta_s+\sin4\beta_s\big)+\nonumber\\
&&\hskip 0pt+ \,
{\textstyle\frac{1}{8}}\Big(\big(39(a^4+b^4)-74a^2b^2\big)\cos4\psi+52ab\big(a^2-b^2\big)\sin4\psi\Big)\big(1+6\cos^2\beta_s+\cos^4\beta_s\big)
\Big]+\nonumber\\
&&\hskip -20pt+ \,
\sin[4(\phi_\xi-\phi_s)]\Big[\Big(ac\big(39a^2-b^2-52c^2\big)\cos\psi+bc\big(a^2-39b^2+52c^2\big)\sin\psi\Big)\sin^3\beta_s+\nonumber\\
&&\hskip 0pt+ \,
\Big({\textstyle\frac{1}{2}}\big(a^2-b^2\big)\big(39(a^2+b^2)-74c^2\big)\sin2\psi-ab\big(13(a^2+b^2)+2c^2\big)\cos2\psi\Big)\sin2\beta_s\sin\beta_s+\nonumber\\
&&\hskip 0pt+
\,\Big(ac\big(13a^2+b^2\big)\cos3\psi+bc\big(a^2+13b^2\big)\sin3\phi\Big)\big(1+3\cos^2\beta_s\big)\sin\beta_s+\nonumber\\
&&\hskip 0pt+ \,
\Big({\textstyle\frac{1}{2}}\big(39(a^4+b^4)-74a^2b^2\big)\sin4\psi-26ab(a^2-b^2)\cos4\psi\Big)\big(1+\cos^2\beta_s\big)\cos\beta_s
\Big]\Big\},
\label{eq:eik-tetra4}
\end{eqnarray}
where one can easily see the harmonic structure of the  gravitational shift due to the hexadecapole STF moment. As before, expression (\ref{eq:eik-tetra4}) may be cast in the familiar harmonic structure:
\begin{eqnarray}
\xi^{[4]}_b(\vec b)&=&kr_g \frac{Q_{\tt t4}}{35840\,b^4_0} \cos[4(\phi_\xi-\phi_s-\phi_{\tt t4})],
\label{eq:eik-tetra4=}
\end{eqnarray}
where the magnitude, $Q_{\tt t4}$, and phase, $\phi_{\tt t4}$,  can readily be read-off directly from (\ref{eq:eik-tetra4}).

Therefore, the gravitational phase shifts introduced by the quadrupole (\ref{eq:eik-tetra2}), octupole (\ref{eq:eik-tetra3}) and hexadecapole (\ref{eq:eik-tetra4}) of the trirectangular tetrahedron obey the same harmonic structure as for other solids where again only two parameters control the magnitude and rotational angle of the resulted caustics. These two parameters depend on the dimensions of the tetrahedron (i.e., $a,b,c$) and its orientation with respect to the observer (i.e., the three Euler angles, $(\phi_s,\beta_s,\psi)$ from Eq.~(\ref{eq:rot})).

\section{Reconstructing the lens from imaging point sources}
\label{sec:appl}

We demonstrated how gravitational lensing by extended compact lenses leads to results that resemble lensing by axisymmetric gravitating bodies whose gravitational field can be represented by a set of zonal harmonics \cite{Turyshev-Toth:2021-STF-moments}. In the case of lenses with arbitrary symmetry, an additional parameter characterizing a rotation appears at each each multipole order. To explore this, we considered lensing by the lower-order mass multipoles of several simple geometric shapes of uniform mass density. This choice, of course, is not intended to imply that there are actual astrophysical lenses that are shaped like a cuboid or a right circular cone. Rather, these cases serve as representative worked examples, showing how, once the tensor moments of inertia of the lens are known, the rest is straightforward: the corresponding lens can be modeled, its PSF and the resulting caustics can be calculated, and the PSF can be convolved with that of an imaging telescope with ease, in a process that is almost mechanical. This simplicity is achieved because the complex three-dimensional structure of the lens is projected onto the thin lens plane. Unfortunately it also implies that we can learn only so much about a particular lens by studying its caustics or the images that it projects, as seen from a single vantage point such as the solar system. That situation would improve if we were able to have observations done by multiple apertures separated by large baselines.

\subsection{Compact formalism}

To investigate if it is possible to distinguish between physically different gravitational lenses by studying the images that they form given a known source, such as a point source, consistently with Eq.~(84) in \cite{Turyshev-Toth:2021-STF-moments}, we introduce a convenient shorthand notation, $\tau_\ell =(2\ell-2)!!\big({t^{+2}_\ell +t^{\times2}_\ell}\big)^\frac{1}{2}/\ell!R^\ell=(2\ell-2)!!\,\sqrt{C_{\ell\ell}^2+S_{\ell\ell}^2}$, where $R$ is the size of the lens and $C_{\ell\ell}$ and $S_{\ell\ell}$ are the appropriate spherical harmonics of its mass distribution (see \cite{Turyshev-Toth:2021-STF-moments} for details). This notation allows us to write (\ref{eq:amp-A+}) in a compact form:
{}
\begin{eqnarray}
A(\vec x) &=&e^{ikr_g\ln 4k^2rr_0}
 \frac{k}{ir}\frac{1}{2\pi}\iint d^2\vec b \,\exp\Big[ik\Big(\frac{1}{2 \tilde r}({\vec b} - \vec x)^2-
 2r_g\Big(\ln kb-\sum_{\ell=2}^\infty
\Big(\frac{R}{b}\Big)^\ell\tau_\ell\cos[\ell(\phi_\xi-\phi_\ell)] \Big)\Big)\Big],
  \label{eq:amp-A++}
\end{eqnarray}
which also allows us to express (\ref{eq:B2-STF}) using a pair of parameters $\tau_\ell$ and $\phi_\ell$ characterizing the contribution of each multipole. An observation of the PSF (given by (\ref{eq:psf=})) or alternatively, an image of a compact (point) source as seen through the lens of an imaging telescope, given by (\ref{eq:BinscER}), allows us to estimate the values of $r_g$ (characterizing the monopole mass), $\tau_\ell$ and $\phi_\ell$ (characterizing contribution of each $\ell$-th multipole).

As a general rule, if an image of a known source is recovered to multipole order $n$ ($2\le\ell\le n$), then by using a suitable numerical optimization method we can recover not just $r_g$, but an additional $2(n-1)$ degrees of freedom in the form of the $\tau_\ell$ and $\phi_\ell$ parameters. This is clearly less information than the full three-dimensional multipole representation of the lens, in the form of $2\ell+1$ degrees of freedom (spherical harmonic coefficients) at each multipole order $\ell$ or the corresponding $2\ell+1$ independent terms in a 3-dimensional STF tensor of rank $\ell$. The question then naturally arises: How much information can be recovered about the mass distribution of a lens? This question can have direct astrophysical significance, as studying lensed images can help reveal information about the mass distribution of the lensing object (e.g., a foreground galaxy or galaxy cluster).

\subsection{Quadrupole lenses}
\label{sec:J2}

Quadrupole lenses yield astroid caustics that may differ in size or phase of rotation, but otherwise they are identical. They have the same fourfold symmetry. The Einstein cross corresponding to a compact, pointlike source, if viewed from a location alongside its optical axis, will exhibit the same symmetry: four spots or arclets of light, equal in shape and intensity, spaced 90$^\circ$ apart. Therefore, if all that can be observed is the effect of the quadrupole moment, nothing is revealed about the internal mass distribution of the lens. (Similar observation on the limited utility of the images produced with quadrupole lenses  for a lens's characterization was made in \cite{Walls-Williams:2018}). This is the case when the shape or orientation of the lens is such that the octupole and higher moments are suppressed, or if the impact parameter is too large for these moments to play a significant role.

\begin{figure}
\centering
\includegraphics[scale=0.91]{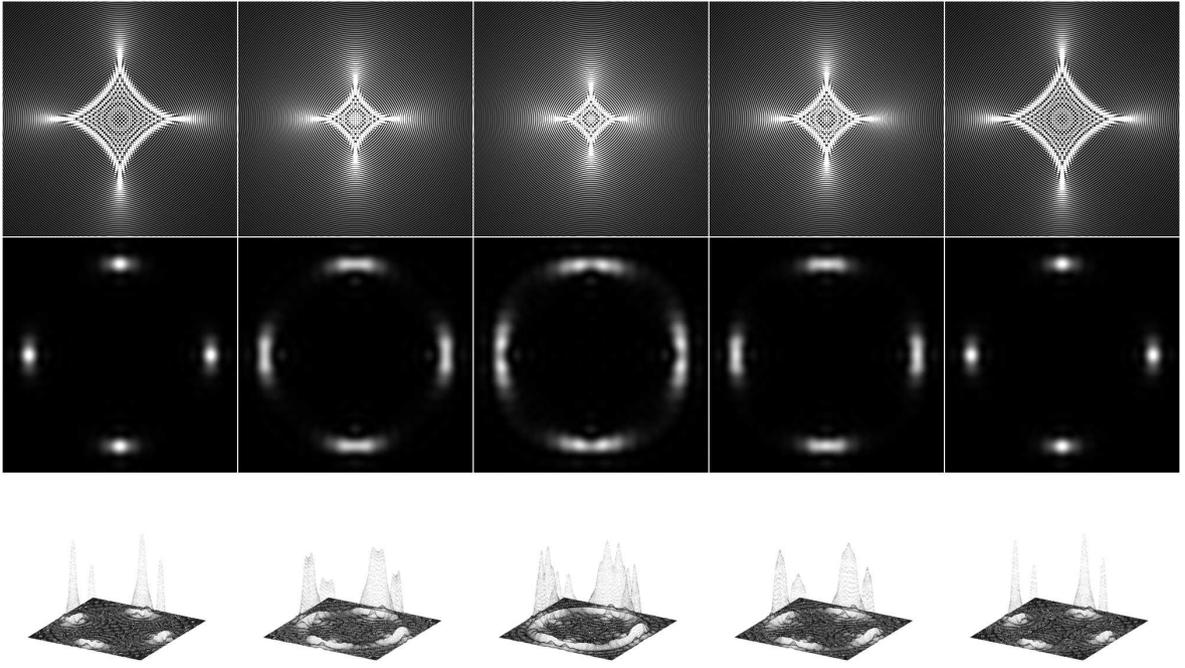}
\vskip -10pt
\caption{\label{fig:montage34}The quadrupole-only ($J_3=J_4=0$) case, discussed in Section~\ref{sec:J2}: PSF (top row), telescopic view (middle row), and a three-dimensional isometric representation (bottom row; to show relative signal amplitudes) of the telescopic view of light projected by the five shapes that we investigated: The ellipsoid, the cuboid, the cylinder, the cone and the tetrahedron. Parameter choices are arbitrary, picked to offer images of comparable size. As only the $J_2$ contribution is shown, the astroids of the PSF differ only in size (the chosen sizes for the shapes were not fine-tuned by us to produce identical sized astroids); they lack distinguishing features. Correspondingly, the telescopic images all share the same basic fourfold symmetry: four identical arcs spaced 90$^\circ$ apart. The smaller the astroid, the wider the arcs; the limiting case when the astroid shrinks to a spit is the full Einstein-ring of the monopole lens.}
\end{figure}

This is demonstrated in Fig.~\ref{fig:montage34}. This figure shows all five of the idealized shapes that we considered, modeled up to the quadrupole moment, using a parametrization that yielded projections of comparable size. As we can see, the images have no distinguishing features. There is no way to tell if an astroid was produced by a cylinder or a tetrahedron. Differences in size can be attributed, e.g., to the magnitude of the quadrupole moment, the size of the impact parameter in relation to the dimensions of the object, or the angle $\beta_s$, revealing nothing about the internal structure of the object in question.

\subsection{Octupole contribution and ``north--south'' asymmetry}
\label{sec:J3}

\begin{figure}
\centering
\includegraphics[scale=0.91]{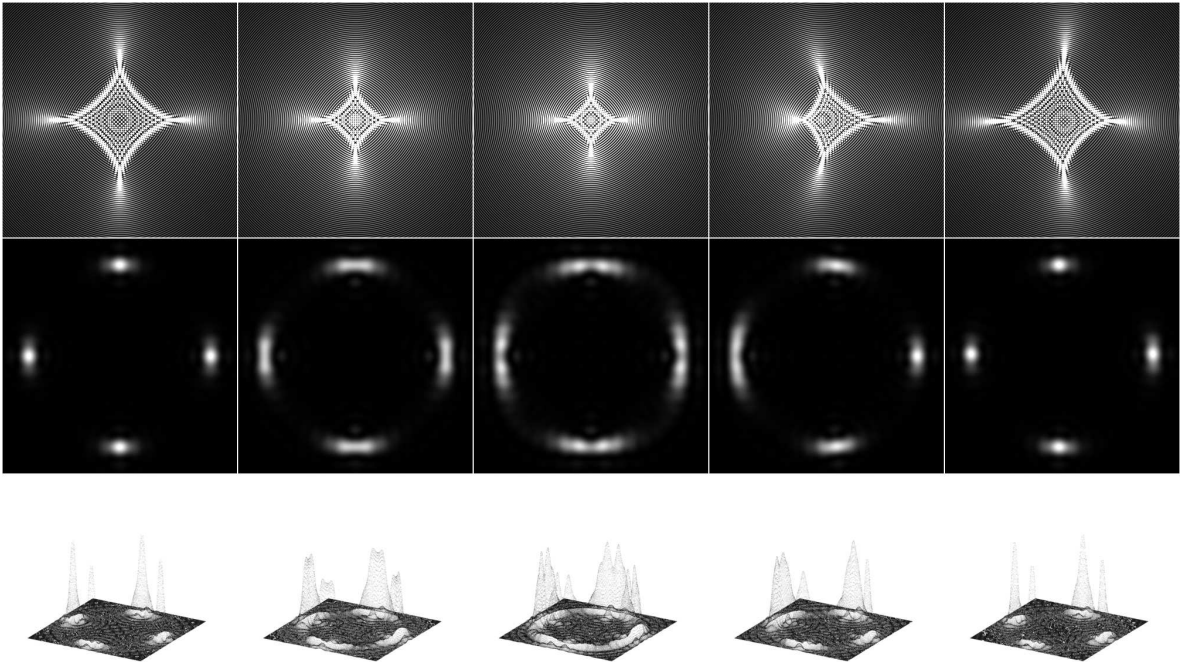}
\vskip -10pt
\caption{\label{fig:montage4}Same as Fig.~\ref{fig:montage34}, but with both the quadrupole and octupole moments shown (as discussed in Section~\ref{sec:J3}), only the hexadecapole suppressed ($J_4=0$). Note that while the ellipsoid, cuboid, and cylinder images remain unchanged, both the cone and the tetrahedron show distinguishing features that break the symmetry of the astroid caustic. The corresponding Einstein crosses in the telescopic view no longer feature four identical arcs of light, but they still retain an axis of symmetry.}
\end{figure}

As we have seen, the cone and tetrahedron lack the ``north--south`` symmetry that characterizes the cuboid, ellipsoid, and cylinder and is indicated by the presence of an octupole moment. This suggests that if the presence of the octupole moment is detected in a gravitational lensing image, we may conclude that the lens lacks ``north--south'' symmetry. In Fig.~\ref{fig:montage4}, we show all five shapes again, up to the octupole moment. The presence of this moment noticeably distorts the astroid caustic. This has an observable impact on the corresponding telescopic images. Whereas quadrupole-only images are characterized by four identical arclets of light, in the presence of the octupole the arclets are unequal in size. However, note that a symmetry is still clearly obvious: in the images representing the cone (fourth column) and tetrahedron (fifth column), the resulting Einstein-crosses, as depicted, both have a horizontal axis of symmetry.
This breaking of the fourfold symmetry of the quadrupole, indicative of the presence of an octupole moment, may be sufficient to distinguish astrophysical lenses that have ``north--south'' symmetry (e.g., a spiral galaxy) from other, irregular lenses. This may also lead to an improved ability to tease out  mass distribution properties of a lens.

\subsection{The hexadecapole and axial symmetry}
\label{sec:J4}

\begin{figure}
\centering
\includegraphics[scale=0.91]{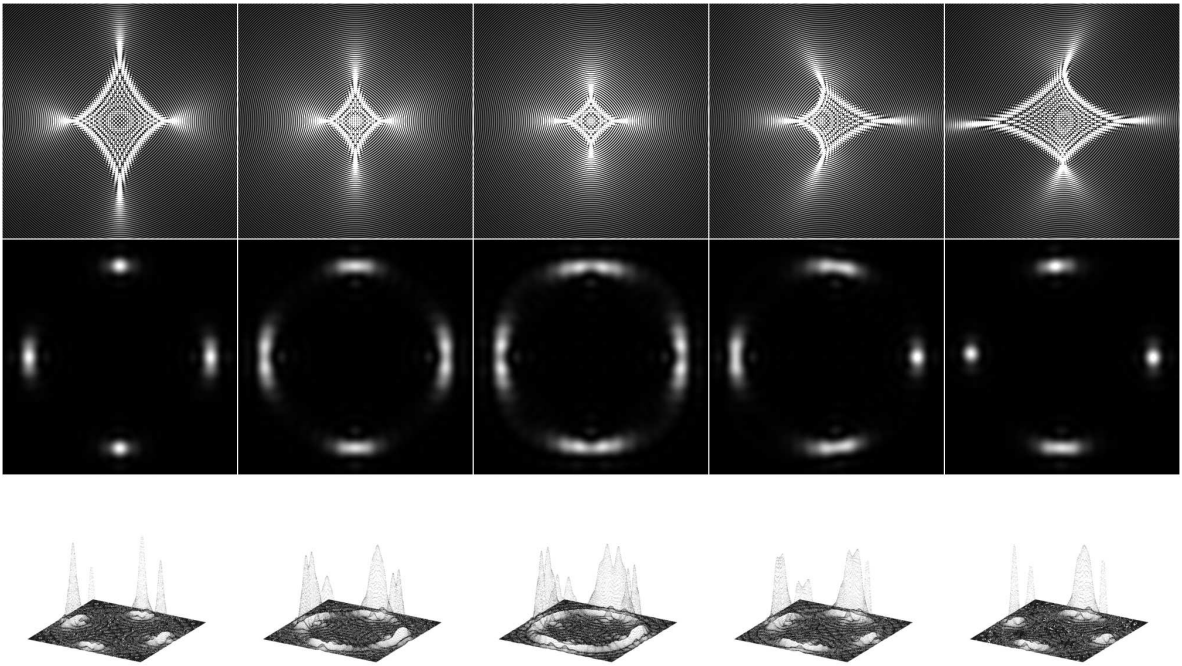}
\vskip -10pt
\caption{\label{fig:montage5}As in Fig.~\ref{fig:montage34} but with all moments up to the hexadecapole present (as discussed in Section~\ref{sec:J4}). The hexadecapole contribution yields a noticeable elongation of the astroid PSF of the ellipsoid and cuboid. These shapes lack axial symmetry. In contrast, the cylinder is axially symmetric and the corresponding PSF and Einstein cross (middle column) retain the fourfold symmetry. The axially symmetric cone (fourth column) yields an Einstein cross that has an axis of symmetry; no such symmetry is present in the Einstein cross of the tetrahedron, which has neither axial nor ``north--south'' symmetry.}
\end{figure}

Finally, we look at lensing by our chosen objects using their moments up to, and including, the hexadecapole moment. The result is depicted in Fig.~\ref{fig:montage5}. What these images reveal is that when the PSF moments can be reconstructed up to the hexadecapole level, important symmetry properties of the lens may be recoverable.

Of the five shapes investigated, only the cylinder (middle column in Fig. ~\ref{fig:montage5}) retains the unbroken fourfold symmetry. This symmetry is retained despite the presence of the hexadecapole moment. While it introduces additional structure to the arclets that form the Einstein cross, the four arclets in the telescopic view remain identical.
The ellipsoid (first column) and cuboid (second column) lack axial symmetry, but have ``north--south'' symmetry. Consequently, the resulting images each have two axes of symmetry. The astroid is elongated, and in the Einstein cross, additional structure appears in the arclets, but the arclets are still spaced 90$^\circ$ apart, and opposing arclets are still identical copies of each other.
The cone (fourth column) shows a more elaborate structure. Its axial symmetry is evident in the shape of the caustic. The resulting telescope image has an axis of symmetry.
However, the trirectangular tetrahedron, which lacks both axial and ``north--south'', yields an irregular, distorted astroid shape. The corresponding Einstein cross has no axis of symmetry and four arclets that are not identical.

\begin{figure}
\centering
\includegraphics[scale=0.91]{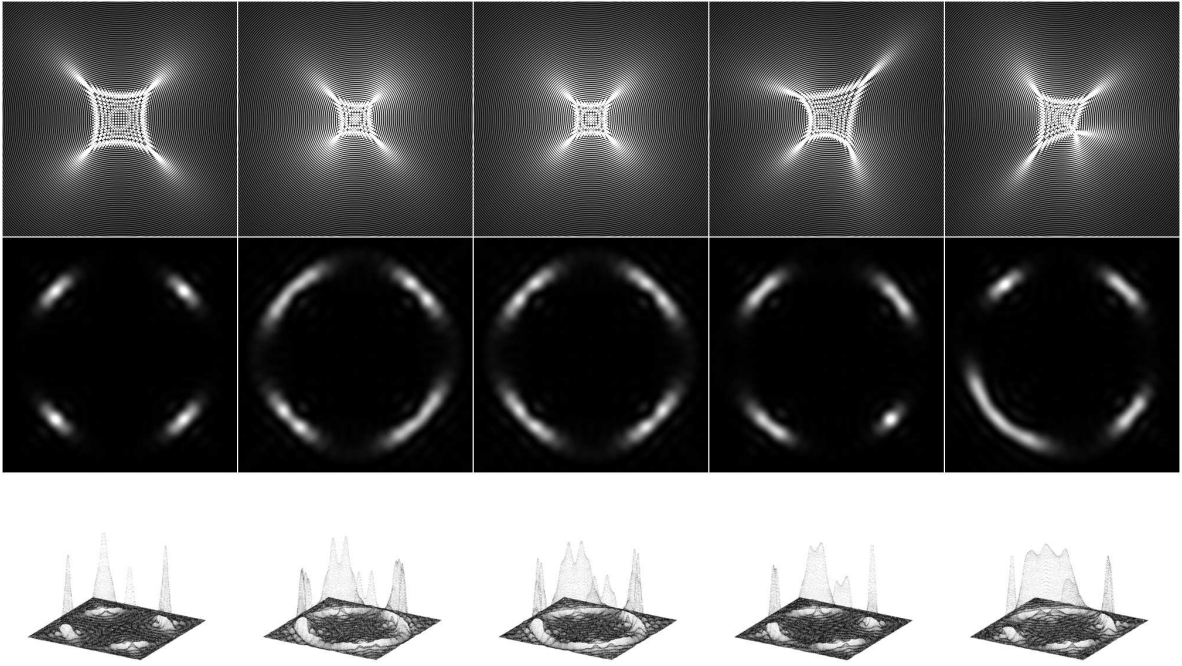}
\vskip -10pt
\caption{\label{fig:montage45}As Fig.~\ref{fig:montage5}, but rotated by $\psi=\pi/4$ in accordance with Eq.~(\ref{eq:rot}). While this changes the orientation and size of the resulting projections and telescopic images, their essential differences remain evident.}
\end{figure}

To what extent are these results artifacts of our parameter choices, in particular the orientation of the shapes that we chose for this demonstration? Fig.~\ref{fig:montage45} depicts the same five shapes, but rotated by $\psi=\pi/4$ in (\ref{eq:rot}). As we can see, this rotation indeed changes both the PSFs and the resulting telescopic images. However, the essential properties that we described above remain. In particular, the cylinder (middle column) retains its fourfold symmetry; the ellipsoid and cuboid still feature two perpendicular axes of symmetry; the cone still has one symmetry axis, whereas the tetrahedron remains the most irregular.

Lastly, we also introduced a rotation by $\beta_{\rm s}=\pi/4$. Once again, the sizes and orientations of the resulting images changed, but they retained the same essential differences in their symmetry properties.

\begin{figure}
\centering
\includegraphics[scale=0.91]{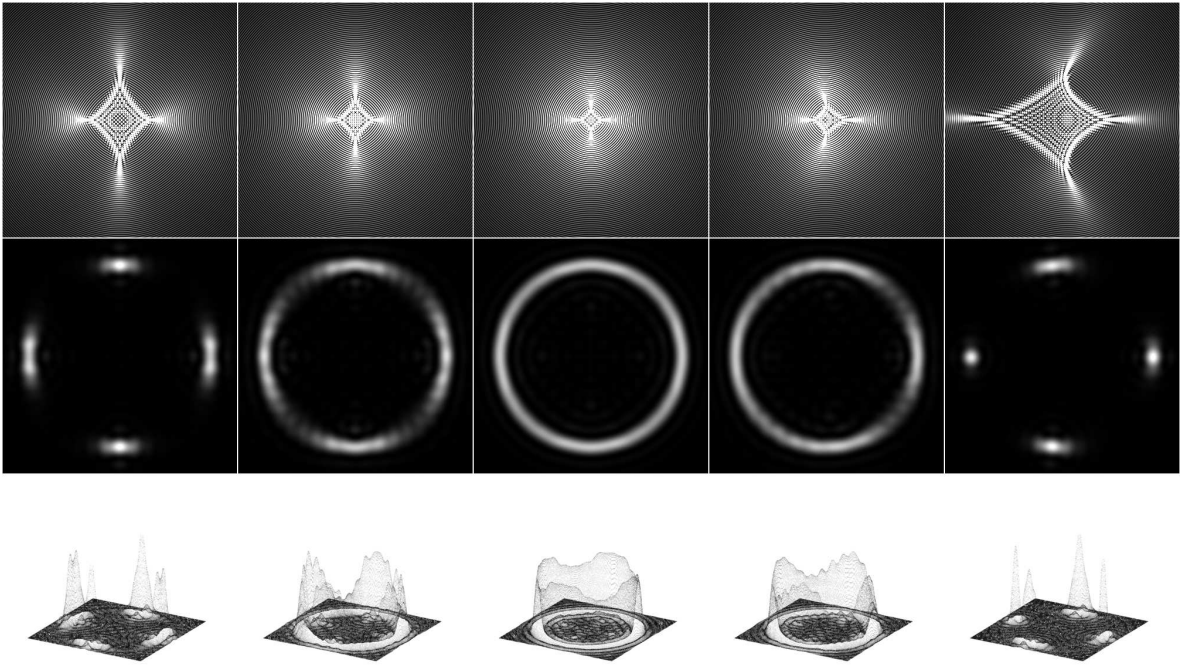}
\vskip -10pt
\caption{\label{fig:montageb45}As in Fig.~\ref{fig:montage45} but rotated this time by $\beta_{\rm s}=\pi/4$.}
\end{figure}

\subsection{Reconstructing the lens from strong lensing observations}

The results presented in the preceding subsections are worth pondering. Investigating only the STF quadrupole moments might have led us to the premature conclusion that due to the degeneracy inherent to the TT projection (see  \cite{Turyshev-Toth:2021-STF-moments}), some shapes may not be distinguishable at all. However, this is not the case, and that is readily demonstrated by the hexadecapole moment of a cuboid and related phase shift (\ref{eq:eik-cuboid4}).
In Table~\ref{tb:shapes} we offer an overview of the various lenses investigated and the resulting properties of the corresponding multipoles. What is remarkable is that despite the inherent loss of information as a result of the TT projection, key symmetry properties of the lens remain observable in the lensed images, so the mass distribution of the lens can at least be partially reconstructed.

\begin{table}
\caption{\label{tb:shapes}Comparison of the STF multipole parameters characterizing the shapes that we investigated. Qualitative distinction of most shapes is possible if the presence or absence of a multipole moment and the corresponding phase angle can be determined unambiguously.}
\begin{center}
\begin{tabular}{l|c|c|c|c|c|}
~           & $J_2\ne 0$ & $J_3\ne 0$ & $\phi_3\ne\phi_2$ & $J_4\ne 0$ & $\phi_4\ne\phi_2$ \\\hline\hline
sphere      &      ~     &      ~     &         ~         &      ~     &         ~         \\
cube        &      ~     &      ~     &         ~         &      x     &         ~         \\
cylinder    &      x     &      ~     &         ~         &      x     &         ~         \\
ellipsoid   &      x     &      ~     &         ~         &      x     &         x         \\
cuboid      &     x     &      ~     &         ~         &      x     &         x         \\
cone        &      x     &      x     &         ~         &      x     &         ~         \\
tetrahedron &      x     &      x     &      x       &      x     &         x      \\\hline
\end{tabular}
\end{center}
\vskip -10pt
\end{table}

This may have real, useful astronomical consequences. The lens associated with strong lensing images may be a foreground object, such as a galaxy. The distribution of visible matter in such a lens may be well known from observation, but much of the lensing mass is presumed to be invisible dark matter. What is its distribution? When visible matter has rotational symmetry, e.g., in a spiral galaxy well modeled by a bulge-and-disk representation, does the corresponding halo of dark matter have the same axial symmetry? Correspondingly, if the lens is an irregular galaxy, should we expect a detectable presence of the $J_3$ moment in lensed images, indicating that the dark matter halo has the same lack of symmetry? Or would dark matter have its own profile, e.g., a spherically symmetric halo \cite{NFW1996} even when visible matter shows irregular distribution?

\section{Discussion and Conclusions}
\label{sec:end}

Descriptions of the gravitational field must deal with the nonlinear nature of the general theory of relativity. Many approximations were developed  for this purpose \cite{MTW,Will_book93}. A weak  field and slow motion approximation (WFSM) \cite{Turyshev-Toth:2013} is often used to describe observations (e.g., see \cite{Will_book93,Turyshev:2012nw,Turyshev-GRACE-FO:2014} and reference therein). Once the WFSM approximation is introduced, the next step is how to describe the source of the gravitational field. The situation is quite simple when dealing with point sources \cite{Turyshev-Toth:2017}; extended sources require a proper description \cite{Turyshev-Toth:2019-extend,Turyshev-Toth:2021-multipoles}.

When dealing with extended objects, their external gravitational potential is often expressed in terms of the spherical harmonics \cite{Turyshev-Toth:2021-multipoles}. When the body has axial symmetry, its gravitational potential is represented purely by zonal harmonics, $J_{2n}, n\in \{ 1,..N\}$. However, even in this case, finding analytical solutions is challenging and only a few of such solutions are known and only for the lowest harmonics, namely for $J_2$ and $J_4$  \cite{Klioner:1991SvA,Will_book93}. An alternative formalism uses STF tensors to represent multipole moments. This representation is mathematically equivalent to spherical harmonics, and allows for the description of light propagation by accounting for multipoles of any order  \cite{Thorne:1980,Blanchet-Damour:1986,Blanchet-Damour:1989,Kopeikin:1997,Mathis-LePoncinLafitte:2007,Soffel-Han:2019}.

We used STF multipole moments to treat gravitational lensing within a wave-optical framework for axisymmetric bodies  \cite{Turyshev-Toth:2021-multipoles, Turyshev-Toth:2021-caustics,Turyshev-Toth:2021-all-regions} as well as bodies with arbitrary mass distributions \cite{Turyshev-Toth:2021-STF-moments}. The STF approach offers a technical advantage by allowing for a closed form solution  while integrating the equations of light propagation in the vicinity of a generic lens. As a result, we were able to develop a wave-optical treatment of gravitational lenses of the most generic structure and internal mass distribution \cite{Turyshev-Toth:2021-STF-moments}. Our powerful approach permits us to not only reconstruct the caustic structure of the lens PSF, but also accurately predict the wavelength-dependent view seen by an observing telescope, including the Einstein-ring of the monopole lens, the Einstein-cross of the quadrupole lens, and more complex cases involving multiple spots and arclets of light that are formed by the lens.

To demonstrate the utility of our approach and to emphasize the physics of the related phenomena, we considered a select set of unusual gravitational lenses in the form of common geometric shapes including the ellipsoid, the cuboid, the cylinder, the right circular cone, and the tetrahedron. We also show how one technically computes the STF multipole mass moments. Specifically, for the least symmetric of these objects, the cone and the tetrahedron, we derived expressions for their octupole moments, and for the cube, cylinder, the right circular cone we computed the hexadecapole moments. We used these results in conjunction with our existing numerical codes to calculate the corresponding PSF and simulated views of lensing as seen through a thin lens imaging telescope.
These modeled objects have different symmetry properties. Some are axially symmetric, some are not. Some have ``north-south'' symmetry, some do not. In that, they resemble various classes of astronomical objects. Although much information is lost when a gravitational lens is studied through the images it projects to an observer at a single vantage point, remarkably, the fundamental symmetry properties of the lens may be recoverable if the presence and phase angle of the $J_3$ and $J_4$ moments can be unambiguously determined. This can help establish better constraints on the dynamics of dark matter halos that surround astronomical objects such as galaxies and galaxy clusters that act as strong gravitational lenses.
We find it remarkable that the mere presence of these moments and phases alone already conveys useful information about the lens.

These results suggest a straightforward approach for recovering the mass distribution of the lens. If a lensed image of a distant source is obtained, a possible path may involve progressively fitting $J_2$; $J_2$, $J_3$ and $\phi_3-\phi_2$; $J_2$, $J_3$, $J_4$, $\phi_3-\phi_2$ and $\phi_4-\phi_2$; and perhaps higher-order moments as well. Even when the uncertainties are substantial, if, for instance, it is possible to determine unambiguously that $J_3\ne 0$ or $\phi_4\ne\phi_2$, this can lead to the conclusion that the mass distribution of the lens lacks ``north-south'' symmetry or axisymmetry. Better fits can of course further constrain the mass model.
At this point, we only considered lensing from a vantage point that is situated on the optical axis: i.e., the axis connecting the center-of-mass of the lens with the point source. Clearly, astrophysical lenses are rarely seen from this special vantage point: when the observation is made from an off-axis position (but still within the caustic region of the multipole PSF) the resulting dipole moment must also be considered, extending the parameter space.

In practice, lens modeling requires both source and deflector models. By treating subcomponents as pointlike, analytic expressions may be obtained for model source flux and position parameters, reducing the parameter search space dimensionality. Image centroids may not correspond to source centroids for non-pointlike subcomponents; a correction for this may be reliably derived within our formalism. In this context, our approach may be used to train and evaluate software for the automatic detection of gravitational arcs and multiple images, as well as for the determination of the mass distribution of the lenses and ultimately to recover cosmological parameters through statistical and geometrical tests, etc.

In the case of the solar gravitational lens (SGL)  \cite{Turyshev-Toth:2017,Turyshev-Toth:2019-extend,Turyshev-Toth:2020-image,Turyshev-Toth:2020-extend,Toth-Turyshev:2020}, all relevant information about the structure of the Sun is well known, allowing for the development of a very realistic model of the extended SGL. This information is now being used in our ongoing analysis of the optical properties of the SGL, especially in the context of a prospective space mission for imaging and spectroscopy of an Earth-like exoplanet \cite{Turyshev-etal:2020-PhaseII}.

Concluding, we emphasize that the results  presented here are new and may be used to study gravitational lensing with a wide range of realistic astrophysical lenses, including many cases that previously could only be modeled by geometric optics and/or ray tracing. This work is ongoing; results, when available, will be published elsewhere.

\section*{Acknowledgements}

This work in part was performed at the Jet Propulsion Laboratory, California Institute of Technology, under a contract with the National Aeronautics and Space Administration.
VTT acknowledges the generous support of Plamen Vasilev and other Patreon patrons.

\section*{Data Availability}

No data was generated and/or analysed to produce this article.



\label{lastpage}

\bibliographystyle{mnras}

\end{document}